\documentclass[twocolumn,floatfix,superscriptaddress,pra,aps,10pt,nofootinbib,longbibliography]{revtex4-2}
\usepackage{times}
\usepackage{amssymb}
\usepackage{color}
\usepackage{amsmath}
\usepackage{amsbsy}
\usepackage{amsthm}
\usepackage{graphicx}
\usepackage{bbm}
\usepackage{bm}
\usepackage{epsfig}
\usepackage{xfrac}
\usepackage{xcolor}
\usepackage{enumerate}
\usepackage{multirow}
\definecolor{myurlcolor}{rgb}{0,0,0.7}
\definecolor{myrefcolor}{rgb}{0.8,0,0}
\usepackage[unicode=true,pdfusetitle, bookmarks=true,bookmarksnumbered=false,bookmarksopen=false, breaklinks=false,pdfborder={0 0 0},backref=false,colorlinks=true, linkcolor=myrefcolor,citecolor=myurlcolor,urlcolor=myurlcolor]{hyperref}

\newcommand{\ket}[1]{\left| {#1} \right\rangle}
\newcommand{\bra}[1]{\left\langle {#1}\right|}

\newcommand{\braket}[2]{\langle #1|#2\rangle}

\renewcommand{\t}[1]{\textrm{#1}}
\newcommand{\tr}[0]{\mathrm{Tr}}

\newcommand{\imag}{\mathrm{Im}}
\newcommand{\diag}{\mathrm{diag}}

\newcommand{\cov}{\Sigma}

\newcommand{\oG}{{\Lambda}}
\newcommand{\bG}{{\boldsymbol \oG}}
\newcommand{\Ch}{{\mathcal E}}

\newcommand{\var}{\theta}

\newcommand{\bvar}{{\boldsymbol{\var}}}
\newcommand{\varn}{{\theta_0}}
\newcommand{\bvarn}{{{\boldsymbol{\var}}_0}}
\newcommand{\B}{{{\var}}}
\newcommand{\bB}{{\boldsymbol{\B}}}

\newcommand{\sep}{\Delta^2\tilde\bvar^{\rm ind}}

\newcommand{\ins}{\text{in}}

\usepackage{braket}
\newcommand{\thmref}[1]{\hyperref[#1]{Theorem~\ref{#1}}}
\newcommand{\lemmaref}[1]{\hyperref[#1]{Lemma~\ref{#1}}}
\newcommand{\figref}[1]{\hyperref[#1]{Fig.~\ref{#1}}}
\newcommand{\figaref}[1]{\hyperref[#1]{Fig.~\ref{#1}a}}
\newcommand{\figbref}[1]{\hyperref[#1]{Fig.~\ref{#1}b}}
\newcommand{\figcref}[1]{\hyperref[#1]{Fig.~\ref{#1}c}}
\renewcommand{\eqref}[1]{\hyperref[#1]{Eq.~(\ref{#1})}}
\newcommand{\eqsref}[2]{\hyperref[#1]{Eqs.~(\ref{#1})-(\ref{#2})}}
\newcommand{\appref}[1]{\hyperref[#1]{Appx.~\ref{#1}}}

\newcommand{\cost}{{\Delta^2\tilde\bvar}}
\renewcommand{\sep}{\cost_{\t{SEP}}}
\newcommand{\sepp}{\cost_{\t{SEP+}}}
\newcommand{\jnt}{\cost_{\t{JNT}}}

\newcommand{\sepcr}{\cost_{\t{SEP}}^{\t{CR}}}
\newcommand{\seppcr}{\cost_{\t{SEP+}}^{\t{CR}}}
\newcommand{\jntcr}{\cost_{\t{JNT}}^{\t{CR}}}

\newcommand{\sepmm}{\cost_{\t{SEP}}^{\t{MM}}}
\newcommand{\seppmm}{\cost_{\t{SEP+}}^{\t{MM}}}
\newcommand{\jntmm}{{\cost_{\t{JNT}}^{\t{MM}}}}

\newcommand{\noon}{\ket{\t{n00n}}}
\newcommand{\siN}{\ket{\t{SIN}}}
\newcommand{\CM}{{\mathcal C}}
\newcommand{\cf}{{\rm cost}}
\newcommand{\limeq}{\simeq}

\usepackage{dsfont}

\begin{document}
\title{Multiparameter quantum metrology in the Heisenberg Limit regime:\\ many repetition scenario vs. full optimization}
\author{Wojciech G{\'{o}}recki}
\affiliation{Faculty of Physics, University of Warsaw, Pasteura 5, 02-093 Warsaw, Poland}
\author{Rafa{\l} Demkowicz-Dobrza{\'n}ski}
\affiliation{Faculty of Physics, University of Warsaw, Pasteura 5, 02-093 Warsaw, Poland}
\begin{abstract}
We discuss the Heisenberg limit in the multiparameter metrology within two different paradigms---the one, where the measurement is repeated many times (so the Cram{\'e}r-Rao bound is guaranteed to be asymptotically saturable) and the second one, where all the resources are allocated into one experimental realization (analyzed with the mimimax approach). We investigate the potential advantage of measuring all the parameter simultaneously compared to estimating them individually, while spending the same total amount of resources.
We show that in general the existence of such an advantage, its magnitude and conditions under which it occurs depends on which of the two paradigms has been chosen. In particular, for the problem of magnetic field sensing using $N$ entangled spin-$1/2$, we show that the predictions based purely on the Cram{\'e}r-Rao formalism may be overly pessimistic in this matter---the minimax approach reveals the superiority of measuring all the parameters jointly whereas the Cram{\'e}r-Rao approach indicates lack of such an advantage.
\end{abstract}

\maketitle

\section{Introduction}

Quantum mechanics opens up new possibilities in metrology, enabling the use of coherence and entanglement to increase measurement precision~\cite{giovannetti2006quantum,Paris2009,giovannetti2011advances,Toth2014,Demkowicz2015,Schnabel2016,degen2017quantum,Pezze2018,Pirandola2018}. The most prominent example of this is the ability to overcome the shot noise-limit linear scaling of the estimation precision with the number of resources $n$ used in measurement (which can be understood as number of photons, total energy, total time, etc.) and obtain a quadratic scaling, the so called Heisenberg Scaling ~\cite{caves1981quantum,holland1993interferometric,lee2002quantum,wineland1992spin,mckenzie2002experimental,bollinger1996optimal,
leibfried2004toward,giovannetti2004quantum,huelga1997improvement,de2005quantum}. Even if presence of decoherence makes the Heisenberg Scaling fragile and virtually impossible to preserve in the asymptotic limit~\cite{escher2011general,demkowicz2014using}, for many models the noise may be completely or partially canceled by applying a proper Quantum Error Correction protocols~\cite{dur2014improved,demkowicz2017adaptive,layden2018ancilla,zhou2018achieving,Gorecki2020}, which allows for the observation of the quadratic precision scaling in certain finite-resource regimes.

If an experiment aimed at estimation a parameter $\theta$ is repeated $k$ times and involves the use of $n$ resources in each repetition, then provided the Heisenberg Scaling holds, the variance of the estimator will scale as:
\begin{equation}
\label{HS}
\Delta^2\tilde\var\propto\frac{1}{k\cdot n^2}.
\end{equation}
If $k$ is sufficiently large, then the problem may be successfully analyzed with the use of the concept of Quantum Fisher Information (QFI) and the related Cram{\'e}r-Rao  (CR) bound (which is proven to be tight in the limit $k\to\infty$). However, as pointed out in ~\cite{hayashi2011,Hall_2012,Pezze_2013,Berry_2015,Hayashi_2018,Gorecki2020pi}, a subtle problem appears if one wants to discuss the best precision achievable when all the available resources $N=n\cdot k$ are used optimally, which we will refer to as the actual Heisenberg Limit.

When inspecting \eqref{HS} it is apparent that, when $N = n \cdot k$ is kept fixed, one should accumulate as much resources as possible in a single repetition of an experiment and therefore increase $n$ at the expense of smaller $k$. Unfortunately, in general it is not clear what is the minimal number of repetition $k$ needed to saturate the CR in practice (which may be different for various models), and hence the sole notion of the QFI does not provide a full understanding of the problem.

This case was broadly discussed for the problem of estimating phase-shift in the interferometer using an $N$-photon state (within
the Bayesian~\cite{luis1996,buzek1999,Berry2000} or minimax ~\cite{hayashi2011} formalism) as well as for general problem of single parameter unitary estimation~\cite{Gorecki2020pi}. It was shown,
that the optimal state is different than the one maximizing the QFI and that the final estimator's variance is $\pi^2$ times larger
then the one resulting from the QFI based analysis)---see \ref{hlsp} for more discussion.

This implies that whenever the Heisenberg Scaling occurs, then in order to discuss the optimal measurement strategy, one needs to strictly define, which paradigm is under consideration.
The one, where all resources $N=n\cdot k$ may be used in the optimal way and accumulated in a single experiment's realization (which in this work we analise within minimax formalism, labeled by MM) or the second one, where the amount of resources used in single trial $n$ is large but finite, and the whole experiment is repeated many times $k$ (analised within Cram{\'e}r-Rao formalism, labeled by CR). In the latter case, the limit $N\to\infty$ corresponds to $k\to \infty, n=\t{const}$. Only such a formulation allows us to apply the general argument about the asymptotical saturability of the CR bounds.


While the issues mentioned above appear now to be completely understood in a single-parameter estimation case, new questions and challenges arise when discussing multi-parameter estimation models~\cite{kolenderski2008,Tsang2011, Berry2013,tsang2016quantum,genoni2013optimal,Liu2017,Nichols2018}. In some situations, a properly designed multiparameter estimation protocol allows to reduce the total error in estimation when compared with a strategy where all the parameters are measured separately in independently
 prepared experiments \cite{albarelli2021probe,ragy2016}. Heisenberg Scaling in multiparameter metrology has been discussed in the literature using both paradigms. Many repetition scenario has been considered in \cite{Yuan2016,baumgratz2016,Gessner2018,Kura2018,ge2018distributed}(using multiparameter quantum CR bound) and~\cite{Holevo1982,matsumoto2002new,demkowicz2020multi,Gorecki2020} (using tighter variants of quantum CR bound), while
 single experiment scenario where the total amount of resources is limited has been analyzed mainly within the Bayesian paradigm
 for models with underlying group symmetry (covariant problems)---SU(2)/U(1)\cite{bagan2000,bagan2001}, SU(2)~\cite{chiribella2004,bagan2004su2,chiribella2005su2,hayashi2006su2}, SO(3)~\cite{hayashi2016fourier}, SU(d)~\cite{kahn2007}.
 The quantitative analysis of the relation between the results obtained within these two paradigms has started to be analyzed only very recently \cite{gorecki2021multiple}.
\begin{table*}[]
\includegraphics[width=2\columnwidth]{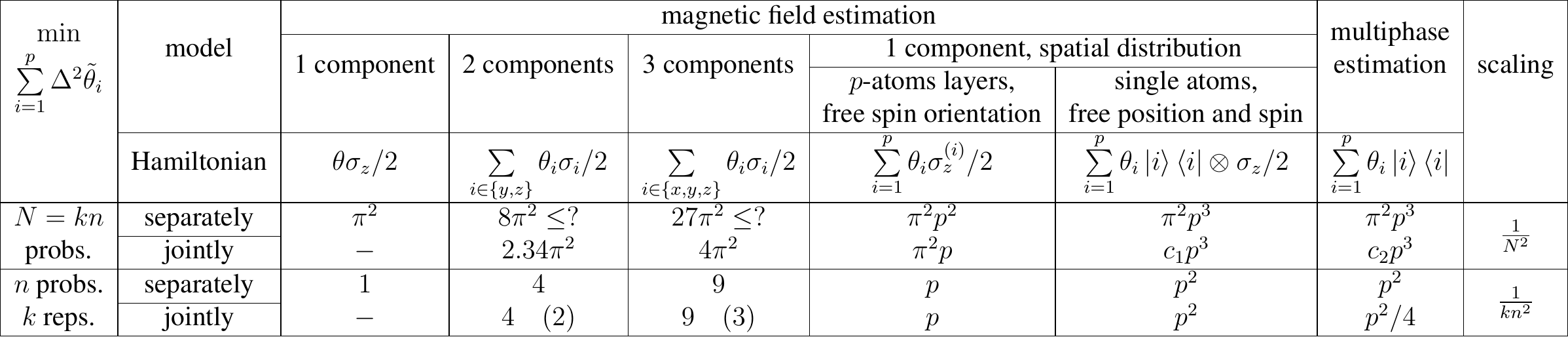}
\caption{Optimal achievable sum of variances of estimated parameters for six different models with unitary evolution governed by the Hamiltonian $\sum\var_i\Lambda_i$. Two paradigms are compared---the one where the channel $e^{i\sum\var_i\Lambda_i}$ is used $N$ times in the optimal way and the second, where it is used $n$ times in each single trial, which is repeated $k$ times. All the values are presented in the limits of large $k$, $n$, $N$ and $p$ (the exact formulas may be found in the main text).
We analyze the optimal strategy when all $p$ parameters are measured separately, as well as the case when they are measured jointly. The asymptotic values and inequalities are written for the parallel strategy case (if adaptiveness help the minimal cost obtainable is written in a bracket). The constants in the two last columns are proven to be $0.63\leq c_1\leq 1$ and  $1.89\leq c_2\leq2$. The $\sigma_z^{(i)}$ in the definition of the Hamiltonian in the middle column is a shortcut for $\openone^{\otimes i-1}\otimes \sigma_z/2\otimes \openone^{\otimes (p-i)}$.
The relation between the costs obtained by different strategies in general depends on the paradigm chosen.}
\label{fig:table}
\end{table*}

The most pressing question is whether a gain can be made by measuring all the parameters simultaneously instead of separately, while consuming the same total amount of resources? That the necessity of splitting these resources between experiments focusing on estimating a given parameter in the separate case, will in general have different consequences in different paradigms~\cite{gorecki2021multiple}. If only the total amount of resources $N$ is restricted, and $p$ parameters are to be estimated, it is rather clear that one needs to spend $\sim N/p$ resources per parameters. Since, we assume the quadratic scaling of precission with the amount of resource used, we may expect the scaling of the sum of variances to be:
\begin{equation}
\label{mmscal}
\sum_{i=1}^p\Delta^2\tilde\var_i\propto p\times \frac{1}{(\sfrac{N}{p})^2}=\frac{p^3}{N^2}.
\end{equation}

In the many repetition scenario, different approaches to analyzing this problem may be found in modern literature. A quite common method is to compare the optimal cost obtainable with applying joint measurement with the one obtained by dividing amount of resources achievable in single trial $n$ between all the parameters, with the total number of repetition of the whole experiment $k$ kept unchanged~\cite{Humphreys2013,goldberg2020,YINGWANG2018250,Yuan2016, baumgratz2016}, which leads to $\sum_{i=1}^p\Delta^2\tilde\var_i\propto p\times \frac{1}{k}\times \frac{1}{(n/p)^2}=\frac{p^3}{kn^2}$. However, one should notice that there is no point in dividing $n$, while we have $k\gg n$ trials at our disposal. It is much more efficient to use $\sim k/p$ trials for each parameter~\cite{yousefjani2017estimating,chen2019,ho2020}. Only such formulation of the problem allows for a fair comparison between measuring parameters jointly vs. separately, and guarantees, that the eventual superiority of the first one comes directly from measuring parameters jointly, not from more efficient resources distribution. As a result we obtain:
\begin{equation}
\label{introcr}
\sum_{i=1}^p\Delta^2\tilde\var_i\propto p\times \frac{1}{\sfrac{k}{p}}\times \frac{1}{n^2}=\frac{p^2}{kn^2},
\end{equation}
which exhibits scaling with $p$ with a different power than in \eqref{mmscal}.

The nontrivial question is, if the sum of variances for the optimal joint measurement will follow a similar scaling---it may turn out that the existence of the advantage depends on paradigm chosen. As recently shown in \cite{gorecki2021multiple}, for the multiphases estimation problem in a multiarm interferometer, this issue does not lead to divergent conclusions---even if the scaling of the total cost with the number of the parameters depends on the paradigm, the potential advantage obtainable by measuring all of the parameters jointly instead of separately
is very similar in both paradigms. In this paper we analyze this issue more generally and show that result is not universal---it may happen that existence or absence  of the advantage indeed depends on the paradigm chosen.

The paper is organized as follows: in section \ref{problem} we remind the basis of the quantum measurement theory, define precisely what we mean by ``measuring parameters separately'' and introduce mathematical formalism useful when analyzing the problem in both paradigms. In section \ref{bound} we derive some general bounds for the achievable precision. Finally, in section \ref{examples} we study multiparameter estimation models representing a variety of magnetic field sensing tasks, check the tightness of introduced bounds and discuss the relation between the results obtained within both paradigms. We also contrast the results of magnetic field sensing models with the multiple arm interferometry case.

The examples are intended to present to the reader in a simple way the diversity and the complicity of the relationship between optimal results obtainable within separate and joint strategies in both discussed paradigms, therefore most of them are easily solvable with basic algebra or are based on analyzing existing results ~\cite{bagan2000,bagan2001,chiribella2004,chiribella2005su2}. However, in \ref{sec:arbitrary} we present also the original results about optimal joint measurement within the minimax approach for spatially distributed magnetic field sensing (which is a specific example with commuting evolution generators). The end results are summarized in table \ref{fig:table} and section \ref{sec:con}.

\section{Problem formulation}
\label{problem}

Let $\Ch_\bvar$ be a quantum channel depending on a vector of unknown parameters $\bvar=[\var_1,...,\var_p]^T$. The aim is to estimate the values of $\bvar$ by sending through the channel some initial state $\Ch_\bvar(\rho_{\rm in})=\rho_{\bvar}$, performing the measurement $\{M_x\}$ (satisfying $\int\t{d}x M_x=\openone$) on the $\rho_{\bvar}$ (leading to probability distribution of the result $x$ given by $p_{\bvar}(x)=\tr(M_x\rho_{\bvar})$) and assigning to outputs of the measurement proper values of estimators $\tilde\var_i(x)$. For such a strategy we define the estimators covariance matrix by:
\begin{equation}
\cov=\int dx\, p_{\bvar}(x)(\tilde\bvar(x)-\bvar)(\tilde\bvar(x)-\bvar)^T,
\end{equation}
and the aim is to minimize its trace, i.e. the sum of squared deviations of the estimator from the true value (for simplicity we will refer to them as variances, implicitly assuming that the estimators will likely be unbiased and their expectation value will coincide with the true value of the parameter) of all parameters:
\begin{equation}
\label{costcov}
\cost:=\sum_{i=1}^p\Delta^2\tilde\var_i=\tr(\cov).
\end{equation}
We will compare the value of the above variance at some reference point $\bvar=\bvar_0$, keeping the condition that the measurement should works well also in some small neighbourhood of this point (two alternative methods how to formalize this condition will be given later).

In the literature, a more general cost function is sometimes considered, where the covariance matrix is additionally multiplied by a positive semidefinite weight matrix under the trace (which makes also non-diagonal elements of the covariant matrix important). Note, however, that such a general case can be reduced to the above after proper reparametrization. Indeed, if for some parametrization $\bvar'$ the cost is given by $\tr(W\cov')$ (with $W\geq 0$), one may take $A$ satisfying $W=A^TA$ and then $\tr(W\cov')=\tr(A^TA\cov')=\tr(A\cov'A^T)=\tr(\cov)$ (where in last step we applied $\bvar=A\bvar'$). Therefore, for simplicity of further formulas we will only consider the cost of the form as given in \eqref{costcov}.

We will consider multiple use of the channel. It may be seen as the action of $N$ gates in parallel on an arbitrary entangled $N$-probe state:
\begin{equation}
\label{parallel}
\rho_{N,\bvar}=\Ch_\bvar^{\otimes N}(\rho_{\rm in}),
\end{equation}
or, even more general, as a general adaptive scheme, where we apply $N$ sequential usage of the channel, where arbitrary large ancilla and arbitrary unitary controls between the actions of the channel are allowed \cite{giovannetti2006quantum, demkowicz2017adaptive, Pirandola2017}:
\begin{equation}
\label{adaptive}
\rho_{N,\bvar}=V_N\circ(\Ch_\bvar\otimes\openone)...V_1\circ(\Ch_\bvar\otimes\openone) (\rho_{\rm in})
\end{equation}
(where $V_i\circ\rho$ is a shortcut for $V_i\rho V_i^\dagger$). Note, that the first one may be simulated by the latter. In such a formulation, the amount of resources corresponds to the number of uses of the channel.

\begin{figure}[t!]
\includegraphics[width=0.5 \textwidth]{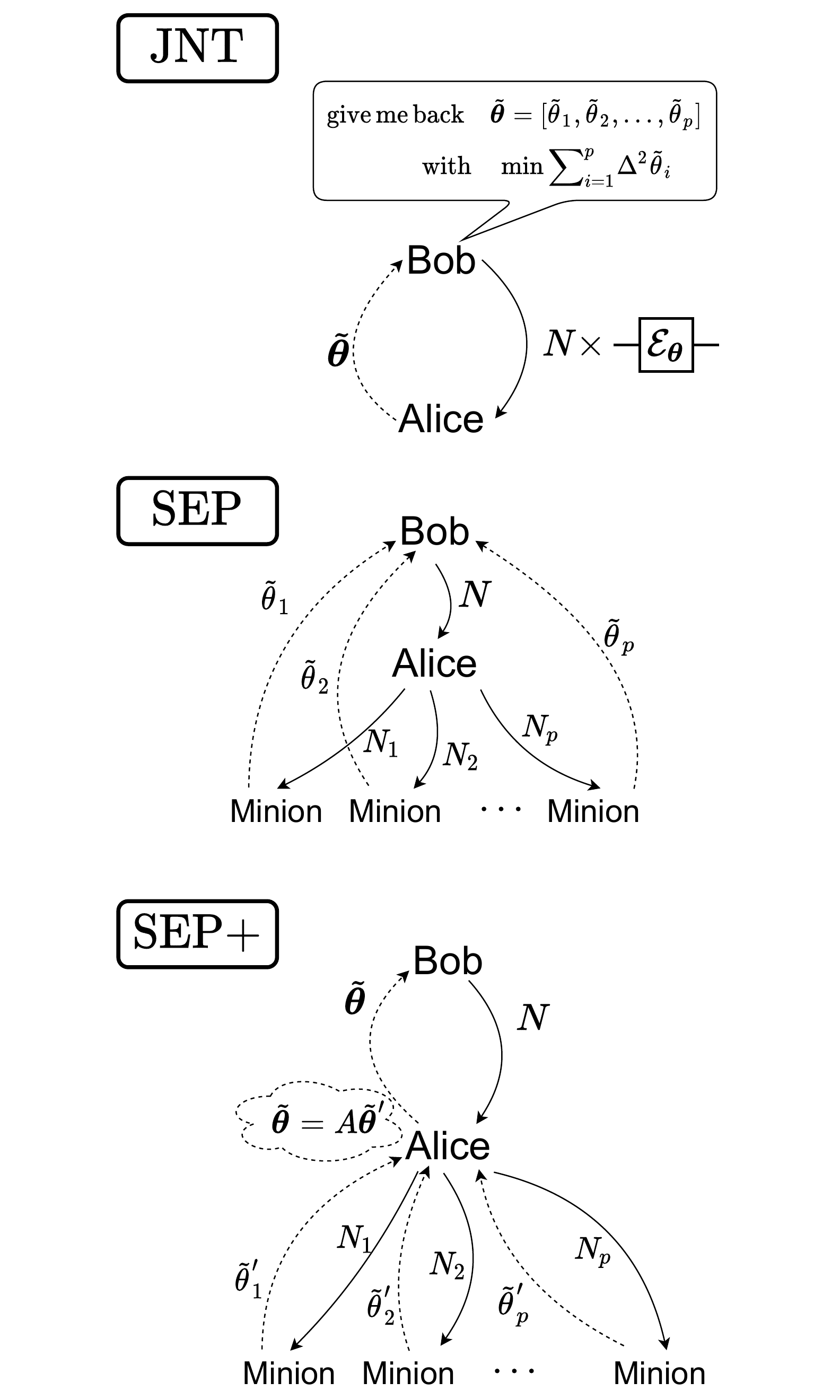}
\caption{Bob sends to Alice $N$ quantum gates that depend on $p$ unknown parameters $\var_i$. Her goal is to send him back their estimated values $\tilde\var_i$, in order to minimize the total quadratic cost. In [JNT] she is allowed perform full optimization and measure all the parameters jointly. In [SEP] she divides the gates between her Minions, where each of them is told to measure a single parameter $\var_i$. In [SEP+] the Minions are allowed to measure mutually linearly independent combinations $\var'_i=[A^{-1}\bvar]_i$ and send the result back to Alice, who reconstructs the initial parameters in post-processing. Note, that [SEP+] may be seen as a special case of [JNT], while [SEP] as a special case of [SEP+].
}
\label{fig:scheme}
\end{figure}

The potential advantage of measuring $p$ parameters jointly vs. separately was extensively discussed in the literature with different approaches, where alternatively the amount of resources used in a single trial~\cite{Humphreys2013,goldberg2020,YINGWANG2018250,Yuan2016, baumgratz2016} or the total number of trials~\cite{yousefjani2017estimating,chen2019,ho2020} was divided between parameters in separate strategy; another analysis free of the resource allocation problem was also performed~\cite{ragy2016,albarelli2021probe} (see \appref{App:incompatibility}  for broader discussion).
In this paper, we would like to focus on the situation, when the same task (i.e. estimation of the set of parameters with a given cost function) is performed with a joint or a separate strategy, when the same amount of resources is used in the end. The problem is schematically shown in \figref{fig:scheme}, where Bob send to Alice $N$ copies of a quantum gate depending on unknown parameters $\bvar$ (where each copy can only be used once) and expect from Alice that she will send him back estimated values of parameters $\tilde\bvar$, in a way that the total cost is minimized.

The Alice may alternatively use $N$ gates to measure all the parameters jointly or separately. In the first case, labeled as JNT, the minimal achievable cost is given by:
\begin{equation}
\label{jnteq}
\jnt=\min_{\substack{N-\t{protocol}}}\tr(\Sigma),
\end{equation}
where by minimization over ``N-protocol'' we understand the minimization over the choice of the initial state $\rho_{\rm in}$, the unitaries $V_i$ acting between $N$ usage of the gate $\Ch_\bvar$, the measurement $\{M_x\}_x$ and the estimator $\tilde\bvar(x)$.

In the second case, labeled as SEP, Alice needs to divide all $N$ gates between her ``Minions'', sending to each of them $N_i$ gates and
ordering to focus on the measurement of a single parameter $\var_i$. Note, that we consider the case when all the parameters have some fixed (unknown) values which do not fluctuate themselves. Hence, in the separate scenario, different parameters are estimated from independent measurements and the resulting covariance matrix will always be diagonal (assuming, that all estimators are unbiased):
\begin{equation}
\Sigma=\t{diag}(\Delta^2\tilde\var_1,...,\Delta^2\tilde\var_p).
\end{equation}
The minimal achievable cost for such a strategy is given by:
\begin{equation}
\label{sepgen}
\sep=\min_{\{N_i\}}\left(\sum_{i=1}^p\min_{\substack{N_i-\t{protocol}}} \Delta^2\tilde{\var}_i\right).
\end{equation}

However, this strategy may by further optimized (while retaining the key feature that each of the Minions measures only a single parameter). Indeed, Alice may demand from them to measure arbitrary linearly independent combinations of original parameters $\var_i'=[A^{-1}\bvar]_i$ (where $A$ is invertible matrix):
\begin{equation}
\Sigma'=\t{diag}(\Delta^2\tilde\var'_1,...,\Delta^2\tilde\var'_p),\quad \cov=A\cov' A^T.
\end{equation}
To distinguish this strategy from the one, where parametrization is fixed, we will label it as SEP+. Note that the resulting covariance matrix is diagonal in $\bvar'$ parametrization, but not necessarily in the initial one. The resulting cost is given by:
\begin{equation}
\sum_{i=1}^p\Delta^2\tilde{\var}_i=\tr(\cov)=\tr(A\cov' A^T)=\sum_{i=1}^p[A^TA]_{ii} \Delta^2\tilde{\var'}_i,
\end{equation}
and therefore:
\begin{equation}
\label{sepgenp}
\sepp=\min_{A,\{N_i\}}\left(\sum_{i=1}^p [A^TA]_{ii} \min_{\substack{N_i-\t{protocol}}} \Delta^2\tilde{\var'}_i\right).
\end{equation}
Note, that for orthogonal transformations $A=O$ where $O^TO=\openone$, no additional term appears in \eqref{sepgenp}. However, it may happen that the optimal separate strategy indeed requires a non-orthogonal transformations (see \appref{App:A} for an example).
Therefore, in general we have:
\begin{equation}
\label{firstin}
\jnt\leq \sepp\leq\sep,
\end{equation}
as for each inequality the right-hand strategy may be seen as the special case of the one on the left side.


We now want to analyze these strategies further using the two paradigms---the one when the constraint is imposed on the number of gates used in a single trial $n$, and where the number of trials $k\gg n$, and the second one where only the total number of gates $N$ is limited.

\subsection{Resources distribution in separate strategies}

Let us start with some general analysis of optimal resource distribution in separate strategies, focussing on the limit of large $N$.
Assume that for each parameter $\var_i$, the minimal variance obtainable with the use of $N_i$ gates (where $\sum_{i=1}^p N_i=N$) scales, in the leading terms, like $N_i^{-\alpha}$:
\begin{equation}
\sep=\min_{\{N_i\}}\sum_{i=1}^p\min_{\substack{N_i-\t{protocol}}} \Delta^2\tilde{\var}_i=\min_{\{N_i\}}\sum_{i=1}^p \frac{c_i}{N_i^{\alpha}}+o(N_i^{-\alpha}),
\end{equation}
where all $c_i$ have finite positive values. Neglecting term $o(N_i^{-\alpha})$, by applying the standard Lagrange multiplayers method, we obtain the optimal resources redistribution to be:
\begin{equation}
N_i=N\frac{c_i^{1/(\alpha+1)}}{\sum_jc_j^{1/(\alpha+1)}},
\end{equation}
which leads to
\begin{equation}
\label{sepopt}
\sep=\frac{1}{N^\alpha}\left(\sum_{i=1}^p c_i ^{1/(\alpha+1)}\right)^{\alpha+1}+o(N^{-\alpha}).
\end{equation}
Further in this paper, we focus only on the asymptotic behavior and for simplicity of the formulas we will omit the term $o(N^{-\alpha})$, using the sign '$\limeq$' instead.
Introducing $\bar{c}_{\alpha}$ for the proper power mean $\bar{c}_{\alpha}=\left(\frac{1}{p}\sum_{i=1}^p c_i ^{1/(\alpha+1)}\right)^{\alpha+1}$ we get:
\begin{equation}
\sep\limeq\frac{\bar{c}_{\alpha}p^{\alpha+1}}{N^\alpha},
\end{equation}
so (for models in which $\bar{c}_{\alpha}$ does not scales with $p$) we see a $p^{\alpha+1}$ scaling of the cost with the number of parameters involved.

Moreover, the optimal SEP and SEP+ strategies  may be bounded from above by:
\begin{equation}
\label{secondin}
\sepp\leq\sep\lesssim p^\alpha \jnt,
\end{equation}
as one may always consider a suboptimal separate strategy, where each of the Minions performs a measurement corresponding to the optimal JNT protocol (with $N/p$ gates), but sends to Alice only the estimated value corresponding to a single parameter $\var_i$.

\subsection{Many repetition scenario ($n$ gates, $k$ trials):\\Cramer-Rao like bounds}

Consider now the situation where the number of gates used in a single trial is restricted to some large but finite number $n$. Then total amount of resources is divided into $k$ trials, satisfying  $N=nk$, and the limit $N\to\infty$ corresponds to $k\to \infty, n=\t{const}$. Therefore, in the context of the discussion from the previous section, the scaling of the cost with the total amount of resources $N$ is always linear here (as it scales linearly with $k$), no matter how it depends on $n$.

Let us briefly remind the foundations of the CR bound approach. The CR bound is based on the idea of local unbiasedness. We consider only estimators satisfying:
\begin{equation}
\label{locun}
\begin{split}
\int dx  p_{\bvarn}(x)\tilde\var_i(x)=\varn_i, \quad
\int dx \frac{d p_\bvar(x)}{d\var_i}\Big|_{\bvar=\bvarn}\tilde\var_j(x)=\delta_{ij}.
\end{split}
\end{equation}
For a given output state $\rho_\bvar$ the covariance matrix is bounded by Quantum CR inequality:
\begin{equation}
\cov\geq\tfrac{1}{k}F^{-1},\quad F_{ij}=\tr\left[\rho_{\bvarn} \tfrac{1}{2}(L_iL_j+L_jL_i)\right],
\end{equation}
where the matrix inequality means that $\cov-\tfrac{1}{k}F^{-1}$ is positive semidefinite and $L_i$ are the symmetric logarithmic derivatives satisfying $\frac{d\rho_\bvar}{d\var_i}\Big|_{\bvar=\bvarn}=\frac{1}{2}(L_i\rho_{\bvarn}+\rho_{\bvarn} L_i)$. This leads to
\begin{equation}
\label{mcr}
\tr(\cov)\geq \tfrac{1}{k}\tr(F^{-1}).
\end{equation}
Since measurements optimal for different parameters might be mutually incompatible, the above inequality is asymptotically saturable for large $k$ if and only if \cite{ragy2016}:
\begin{equation}
\label{imag}
{\rm Im}\tr(\rho_{\bvarn}L_iL_j)=0.
\end{equation}
More precisely, if \eqref{imag} is satisfied and additionally $\rho_{\bvar}$ is pure, then there exists a local measurement (performed on a single copy of $\rho_{\bvar}$) and an estimator depending on $k$ measurement results $\tilde\bvar(x_1,x_2,...,x_k)$, which is asymptotically unbiased and saturate \eqref{mcr}. However, if $\rho_{\bvar}$ is the mixed state, in may in general happen that a collective measurement on all $k$ copies of the state $\rho_{\bvar}^{\otimes k}$ is required in order to saturate the CR bound \cite{Holevo1982,demkowicz2020multi}. Therefore, the minimal achievable cost obtainable with $k$ trials, where in each trial $n$ gates are used, may be bounded from below by the RHS of \eqref{mcr} minimized over all feasible output states $\rho_{n,\bvar}$ (i.e. states that can be obtained when optimizing the protocol over $\rho_{\t{in}}$ and all $V_1,...V_n$ in \eqref{adaptive}):
\begin{equation}
\label{crjnt}
\jntcr\geq\tfrac{1}{k}\min_{\rho_{n,\bvar}}\tr(F^{-1}),
\end{equation}
where the inequality is tight iff \eqref{imag} holds---in fact, in all examples discussed in this paper this will be the case.


Now let us consider a separate strategy and let $\rho^{i}_{n,\bvar}$ be the output state designed to estimate the value of parameter $\var_i$. Note, that in principle it may by inefficient or even impossible to fully isolate the dependence of the state on the remaining parameters $\var_{j\neq i}$---their variations may affect the measurement results, and they cannot be omitted in the analysis of the optimal separate protocol.
That means that the remaining parameters should in general be treated as nuisance parameters \cite{suzuki2020nuisance,suzuki2020quantum}.
Denoting by $F_i$ the QFI matrix corresponding to $\rho^{i}_{n,\bvar}$, the minimal variance of estimating parameter $\var_i$
using $k_i$ repetitions of an experiment is bounded by:
\begin{equation}
\label{eq:singlenuisance}
\Delta^2\tilde\var_i\geq \frac{1}{k_i}[F_{i}^{-1}]_{ii}.
\end{equation}
Strictly speaking, it is not necessary for the whole matrix $F_{i}$ to be invertible---it is enough if $\lim_{\epsilon\to 0^+}[(F_i+\epsilon\openone)^{-1}]_{ii}$ converges to a finite value. For example, if $F_{i}$ has a block-diagonal structure, it is enough if only the block containing $[ \cdot ]_{ii}$ element is invertible.

Note that in general $[F_{i}^{-1}]_{ii}\geq [F_{i}]_{ii}^{-1}$, where the RHS of the inequality corresponds to a direct
application of the single-parameter estimation theorem---neglecting the role of nuisance parameters.
Moreover, \eqref{eq:singlenuisance} is always saturable for large $k_i$ if collective measurements allowed \cite[Section 2.7]{demkowicz2020multi}, and therefore, \eqref{sepopt} (with $\alpha=1$) takes the form:
\begin{equation}
\sepcr\limeq\frac{1}{k}\left(\sum_{i=1}^p\sqrt{\min_{\rho^i_{n,\bvar}}[F_i^{-1}]_{ii}}\right)^2.
\end{equation}
After optimization over the parametrization is performed we get:
\begin{equation}
\label{crsep}
\seppcr\limeq\frac{1}{k}\min_{A}\left(\sum_{i=1}^p\sqrt{[A^TA]_{ii}\min_{\rho^i_{n,\bvar}}[A^{-1}F_i^{-1}A^{-1T}]_{ii}}\right)^2,
\end{equation}
where the rule for the transformation of the QFI matrix $F_i'=A^{T}F_iA$ has been used.

Using \eqref{firstin} and \eqref{secondin} we can write:
\begin{equation}
\jntcr\leq\seppcr\leq\sepcr\lesssim p\cdot\jntcr.
\end{equation}

\subsection{Heisenberg Limit (total $N$ gates):\\asymptotic local minimax bound}

The methods discussed in the previous section cannot be used in a situation, where only the total amount of gates $N$ is constrained,
since we can not invoke the general CR saturability arguments which require a many repetition scenario. In the single parameter case the problem has been discussed within the MM~\cite{hayashi2011} and the Bayesian~\cite{Gorecki2020pi} approach.

In this paper we will follow the first one, as it is conceptually and technically simpler to apply in the case of
estimating values of a parameter in a vicinity of a single point. Let us start by briefly reminding the idea of
local asymptotic MM approach applied to single parameter estimation \cite{hajek1972local,hayashi2011}.

Instead of invoking the property of local unbiasedness (as is done in the CR based approach), we assume that the true value of the parameter lies in some finite-size neighborhood of $\varn$, named $\Theta(\var_0,\delta)=[\varn-\delta/2,\varn+\delta/2]$.
  Then, we consider a strategy, which minimizes the cost in the most pessimistic scenario (we always choose the point in $\Theta$, where the strategy works the worst). Here, unlike in the previously discussed approach, only a single realization of the measurement is considered, and any measurement outcome is directly related with a given value of the estimator $\tilde\var$. Due to this fact, and for the simplicity of notation, we may label the measurement's outcomes by $\tilde\var$, so that the formula for the MM bound takes the form:
\begin{equation}
\label{mmgen}
\inf_{\rho_\var, M_{\tilde\var}}\sup_{\var\in\Theta(\var_0,\delta)}\int d\tilde\var \tr(M_{\tilde\var}\rho_\var)(\tilde\var-\var)^2.
\end{equation}
This value, however, depends on the size of $\Theta$. In order to get rid of this dependence, and hence be able to compare the results
with the CR based approach (which is effectively a single-point estimation approach), we do the following construction.
Let $\{\rho_{N,\var},M_{N,\tilde\var}\}$ be a sequence of output states (for an $N$-gate protocol) and the corresponding measurements. Then, assuming that for a large $N$ the corresponding cost scales like $1/N^{\alpha}$ (where $\alpha=1$ corresponds to the standard scaling, while $\alpha=2$ to the Heisenberg Scaling), define~\cite[Section 5]{hayashi2011}:
 \begin{multline}
 \label{minimax}
\CM^{(\alpha)}[\{\rho_{N,\var},M_{N,\tilde\var}\}]=\\\lim_{\delta\to 0}\lim_{N\to\infty}N^{\alpha}\sup_{\var\in\Theta(\var_0,\delta)}\int \t{d}\tilde\var \tr(M_{N,\tilde\var}\rho_{N,\var})(\tilde\var-\var)^2.
\end{multline}
Note, that the order of taking the limits matters (as for the opposite order the trivial constant estimator pointing $\var_0$ independent on the measurement results would lead to a zero cost). Intuitively, such an order approximates a situation, where one consider a $\delta$-independent measurement strategy and for each $\delta$ checks its validity only for $N$ which is much larger than the inverse of $\delta$. Finally, taking the limit $\delta\to 0$ makes the results independent of the fact that the estimation around different points in $\Theta$ may be in principle harder than around $\var_0$ (i.e. even the value of the maximal QFI may depend on the value of $\var$, as for example in ~\cite{baumgratz2016}).

It was shown in \cite[Section 5]{hayashi2011} that if for a given channel the Heisenberg Scaling is not achievable ($\alpha=1$), then indeed
\begin{equation}
\inf_{\{\rho_{N,\var},M_{N,\tilde\var}\}}\CM^{(1)}=\lim_{N\to\infty}\inf\limits_{\rho_{N,\bvar}}\frac{N}{F}.
\end{equation}
Therefore the MM (in the limit $\delta \rightarrow 0$) and the CR based approaches return consistent results. However, if the channel estimation problem admits the Heisenberg Scaling ($\alpha=2$) then
\begin{equation}
\inf_{\{\rho_{N,\var},M_{N,\tilde\var}\}}\CM^{(2)}\geq\lim_{N\to\infty}\inf\limits_{\rho_{N,\bvar}}\frac{N^2}{F},
\end{equation}
where the inequality is not tight in general (see the next section).
In this paper we are focusing only on the case $\alpha=2$, and hence in what follows for a more compact notation we will drop the
upper index.

For a general multiparameter estimation problem, let us define
 $\Theta(\bvar_0,\delta)=[\var_{0,1}-\delta/2,\var_{0,1}+\delta/2]\times...\times[\var_{0,p}-\delta/2,\var_{0,p}+\delta/2]$ and
\begin{multline}
\label{mmformal}
\CM_i[\{\rho^i_{N_i,\bvar},M_{N_i,\tilde\var_i}\}]\\
=\lim_{\delta\to 0}\lim_{N_i\to\infty}N_i^{2}\sup_{\bvar\in\Theta(\bvar_0,\delta)}\int d\tilde\var_i \tr(M_{N_i,\tilde\var_i}\rho^i_{N_i,\bvar})(\tilde\var_i-\var_i)^2.
\end{multline}
Then, for a large $N$, \eqref{sepopt} (with $\alpha=2$) up to the leading term in $N$ takes the form:
\begin{equation}
\sepmm\limeq\frac{1}{N^2}\left(\sum_{i=1}^p\sqrt[3]{\inf_{\{\rho_{N_i,\bvar},M_{N_i,\tilde\var_i}\}}\CM_i}\right)^3.
\end{equation}
After the optimization over reparametrizations it reads:
\begin{equation}
\label{mmsep}
\seppmm\limeq\frac{1}{N^2}\min_{A}\left(\sum_{i=1}^p\sqrt[3]{[A^TA]_{ii}\inf_{\{\rho_{N_i,\bvar'},M_{N,\tilde\var'_i}\}}\CM'_i}\right)^3,
\end{equation}
where $\CM'_i$ is given by \eqref{mmformal}, after making the substitution $\var_i\to\var_i'=[A^{-1}\bvar]_i$. Analogously, for the joint estimation case:
\begin{multline}
\label{mmlocal}
\CM_{\t{JNT}}[\{\rho_{N,\bvar},M_{N,\tilde\bvar}\}]\\
=\lim_{\delta\to 0}\lim_{N\to\infty}N^{2}\sup_{\var\in\Theta(\bvar_0,\delta)}\int d\tilde\bvar \tr(M_{N,\tilde\bvar}\rho_{N,\bvar})\tr(\cov)
\end{multline}
and
\begin{equation}
\label{mmjnt}
\jntmm\limeq\frac{1}{N^2}\inf_{\{\rho_{N,\bvar},M_{N,\tilde\bvar}\}}\CM_{\t{JNT}}.
\end{equation}
Finally, using \eqref{firstin} and \eqref{secondin} we can write:
\begin{equation}
\jntmm\leq\seppmm\leq\sepmm\lesssim p^2\cdot\jntmm.
\end{equation}

\section{Heisenberg Limit bound}
\label{bound}
Further on we will be interested in the models where Heisenberg Scaling occurs for all the estimated parameter. This is the case for noiseless unitary evolution, where the parameters enter into the evolution as multipliers of the evolution generators:
\begin{equation}
\label{multi}
U_\bvar=e^{i \bvar\cdot\bG},
\end{equation}
with $\bG=[\oG_1,...,\oG_p]^T$ where all $\oG_i$ are mutually linearly independent.

Note, that in presence of noise, Heisenberg Scaling may not be achieved in general \cite{escher2011general, demkowicz2012elusive}. Still, for certain noise models, a proper quantum error correction protocol may be used to  isolate the part of signal which is undisturbed by the noise and effectively obtain a purely unitary evolution \cite{demkowicz2017adaptive,zhou2018achieving,Gorecki2020, Zhou2021} of the form \eqref{multi}. Hence, our discussion here, will be relevant for such models as well.

\subsection{Single parameter staurable lower bound}
\label{hlsp}
As a reference point for further considerations, let us recall a paradigmatic estimation model---a single phase estimation problem in a two arm interferometer. In this case, the parameter encoding channel acts on the two-mode single photon states space spanned by $\{\ket{0},\ket{1}\}$, and is represented by:
\begin{equation}
U_\var=e^{i\var\ket{1}\bra{1}}.
\end{equation}
For simplicity, let us focus on parallel strategies \eqref{parallel} first. The output state $\ket{\psi^n_\var}=U^{\otimes n}_\var\ket{\psi_{\ins}}$, which maximizes the QFI is the famous $\noon$ state:
\begin{equation}
\label{noon}
\ket{\psi^n_\var}=\frac{1}{\sqrt{2}}(\ket{0}^{\otimes n}+e^{in\var}\ket{1}^{\otimes n}),
\end{equation}
for which $F=n^2$
and CR the bound may be saturated with protective measurement onto states $\ket{\pm}=\frac{1}{\sqrt{2}}(\ket{0}^{\otimes n}\pm i\ket{1}^{\otimes n})$. In fact, it may be also saturated by a standard photon counting measurement and an estimator based on the parity of detected number of photons~\cite{chiruvelli2011parity}.

Note, however, that the state \eqref{noon} is unable to distinguish between the phases that differ by a multiple of $2\pi/n$~\cite[Section 5]{hayashi2011}. Therefore it only allows for estimation of the parameter in a small region of $[\var_0-\pi/n,\var_0+\pi/n]$. While many repetition scenario is under consideration, this issue does not generate a serious problem, as even starting with the unknown phase, for $k\gg n$ one may always spend the first $\sqrt{k}$ trials to find such small region (for example by using the product states $\frac{1}{\sqrt{2}^n}(\ket{0}+\ket{1})^{\otimes n}$ in each repetition) and next in remaining $k-\sqrt{k}$ trials use $\noon$ states to finally achieve precision $1/kn^2$ (up to the leading term in $k$). However, it is clear that an analogous single $\ket{N00N}$ state cannot be used to obtain the fundamental Heisenberg Limit---if inserted into \eqref{minimax} it would lead to $\CM=+\infty$.

Calculation of the minimal obtainable value of $\CM$ by a direct minimization of \eqref{minimax} is a hard problem. Fortunately, the task may be significantly simplified by taking into account a symmetry of the problem.

First note, that for such a channel the two phases which differ by a factor $2\pi$ should be regarded as equivalent.
Therefore, we consider a periodic cost function of the form:
\begin{equation}
\cf(\var,\tilde\var)=4\sin^2\left(\tfrac{\var-\tilde\var}{2}\right),
\end{equation}
which reflects this property whereas for small difference may be well approximated by $\approx (\var-\tilde\var)^2$.

Next, as argued in \cite{hayashi2011}, the local asymptotic MM cost \eqref{minimax} for this problem is exactly the same as the
minimal obtainable cost for a completely unknown phase $\theta\in\Theta=[0,2\pi)$ multiplied by $N^2$ (in the limit $N\to\infty$).
Intuitively, the reason for this is that for any finite $\delta$, when we start from a completely unknown phase we always spend at the beginning $~\sqrt{N}$ gates to discriminate the region of size $\delta$, where the true value of $\var$ lies (with the probability of the error decreasing as exponentially fast with $N$), and use the remaining $N-\sqrt{N}$ to estimate the value inside this region.

Note, that while at a first glance the above construction seems to require adaptivness (as in the second step we use the information from the first one), in may in fact be performed also within the parallel scheme (see detais in \appref{App:cost}). See also \cite{Imai2009,Gorecki2020pi,gorecki2021multiple} for further discussion about optimization of region discrimination in the first part of the above construction, as well as a general discussion about this bound for finite $N$.

The state which is optimal for measuring a completely unknown phase is the $\siN$ state \cite{luis1996,buzek1999,Berry2000}:
\begin{equation}
\label{sin}
\ket{\psi^N_{\bvar}}=\sum_{m=0}^Ne^{im\var}\frac{\sqrt{2}}{\sqrt{N+2}}\sin\left(\frac{(m+1)\pi}{N+2}\right)\ket{N-m}_0\ket{m}_1,
\end{equation}
where $\ket{N-m}_0\ket{m}_1$ is a fully symmetric state with $m$ photons in the sensing arm $\ket{1}$ and $N-m$ is the reference arm $\ket{0}$. The corresponding mean cost obtainable with applying covariant measurement is:
\begin{equation}
\forall_{\var}\int d\tilde\var \tr(M_{N,\tilde\var}\rho_{N,\var})\cf(\var,\tilde\var)=2\left(1-\cos\left(\tfrac{\pi}{N+2}\right)\right)
\end{equation}
and therefore the constant which multiplies the leading term $1/N^2$ equals:
\begin{equation}
\inf_{\{\rho_{N,\bvar},M_{N,\tilde\var}\}}\CM=\lim_{N\to\infty}N^22\left(1-\cos\left(\tfrac{\pi}{N+2}\right)\right)=\pi^2.
\end{equation}

The natural question arises: is there a simple interpretation of this $\pi^2$ factor discrepancy between the minimal achievable variance and the inverse of maximal Fisher information? In fact, one may indeed use the $\noon$ states even
when estimating a completely unknown phase, provided one divides all the available resources $N$ into $M$ subsets, each of size $m_i$ and then use $M$ states of the form $\frac{1}{\sqrt{2}}(\ket{0}^{\otimes m_i}+\ket{1}^{\otimes m_i})$ for estimation. It was shown numerically, that for the optimal distribution of $m_i$, the overhead factor $\pi^2$ is indeed recovered \cite{Kaftal2014,berry2009}.
Some suboptimal strategies have also been demonstrated experimentally, revealing only a slightly bigger variance \cite{higgins2007entanglement,higgins2009demonstrating}

The above result may be directly generalized for an arbitrary quantum channel of the form $U_\var=e^{i\var\Lambda}$ \cite{Gorecki2020pi}.
Denoting  by $\lambda[\oG]$ the difference between the maximal and the minimal eigenvalues of the operator $\oG$, for optimal usage of $N$ quantum gates we have:
\begin{equation}
\label{mmpi}
\inf_{\{\rho_{N,\bvar},M_{N,\tilde\var}\}}\CM=\frac{\pi^2}{\lambda^2[\oG]}\Rightarrow \Delta^2\tilde\var^{\t{MM}}\limeq \frac{\pi^2}{N^2\lambda^2[\oG]}.
\end{equation}
For comparision, in the scenario, where $k\to\infty$ repetitions is considered (with usage $n$ gates in each of them), minimal obtainable cost is given by:
\begin{equation}
\label{cr}
\max_{\rho_{n,\var}} F=n^2\lambda^2\Rightarrow \Delta^2\tilde\var^{\t{CR}}\limeq \frac{1}{kn^2\lambda^2[\oG]}.
\end{equation}
In both scenarios, this optimal precision is obtainable already in the parallel scheme and it cannot be beaten by any adaptive protocol
that involves additional action of $V_i$ operations in between~\cite{Gorecki2020pi}.

\subsection{Multi-parameter unitary estimation lower bound}

Consider a general problem of local unitary channel estimation:
\begin{equation}
\label{multi2}
U_\bvar=e^{i \bvar\cdot\bG},
\end{equation}
with linearly independent generators $\oG_i$ acting on a $d$ dimensional space. First note that for such a formulated problem the Heisenberg Scaling is indeed achievable. This was shown for the most general $SU(d)$ estimation problem for any $d$ using a parallel scheme---in both many repetition scenario~\cite{Imai_2007} (where the exact fundamental bound has been derived and proven to be saturable if entanglement with ancilla allowed) and single repetition scenario~\cite{kahn2007} (where optimal scaling has been proven to be $\propto 1/N^2$ with an exemplary state satisfying this scaling).
As for any finite dimensional space, the problem states in \eqref{multi2} may be seen as estimation of some subset of $SU(d)$ generators, the statement is proven.

Let us first argue, that from the point of view of estimation of any single parameter $\var_i$,
the existence of an additional part of the generator $\sum_{j\neq i}\var_j\oG_j$ cannot help in estimation, i.e. it cannot decrease the minimal achievable cost. Note, that a single gate $U_\bvar$ (where all generator acts jointly) may be arbitrarily well approximated (for a sufficiently large $l$) by:
\begin{equation}
U_\bvar=e^{i\bvar\cdot\bG}\overset{l\gg 1}{\approx} \left(e^{i\var_i \frac{\oG_i}{l}}e^{i\sum_{j\neq i}\var_j\frac{\oG_j}{l}}\right)^l
\end{equation}
(more precisely, the above approximation is exact in the limit $l\to\infty$ due to the Trotter formula~\cite{COHEN198255}). Therefore an $N$-fold action of $U_\bvar$ may be seen as $l\cdot N$ action of $U_{\var_i}=e^{i\var_i \frac{\oG_i}{l}}$, with unitary controls $V=e^{i\sum_{j\neq i}\var_j\frac{\oG_j}{l}}$ in between. After such a procedure the product 
$\lambda[\Lambda_i/l]\cdot (lN)=\lambda[\Lambda_i]N$ remains unchanged. Therefore, both the asymptotic value of the bound and the rate of its convergence remain the same. Consequently, for each $\var_i$ we have:
\begin{equation}
\inf_{\{\rho^i_{N_i,\bvar},M_{N_i,\tilde\var}\}} \CM_i\geq \frac{\pi^2}{\lambda^2[\Lambda_i]}.
\end{equation}
Note, however, that the saturability of the bound is not guaranteed.
Let us now bound from below the minimal achievable cost by assuming the most optimistic scenario---the one where not only the existence of the other part of the Hamiltonian $\sum_{j\neq i}\var_j\oG_j$ does not disturb the sensing of $\var_i$, but also that there exist a single input state and a single measurement which are simultaneously optimal for sensing of all the parameters.  Then, for any finite $N$ we have:
\begin{equation}
\label{mbounda}
\min_{\rho_{N,\bvar},M_{N,\tilde\bvar}}\tr(\cov)=\min_{\rho_{N,\bvar},M_{N,\tilde\bvar}}\sum_{i=1}^p\Delta^2\tilde\var_i
\geq
\sum_{i=1}^p\min_{\rho^{i}_{N,\bvar},M_{N,\tilde\var_i}}\Delta^2\tilde\var_i
\end{equation}
and therefore, taking the asymptotic limit $N\to\infty$, we can  write:
\begin{equation}
\label{mmmm}
\inf_{\{\rho_{N,\bvar},M_{N,\tilde\bvar}\}} \CM_{\t{JNT}}\geq \sum_{i=1}^p\inf_{\{\rho^i_{N,\bvar},M_{N,\var_i}\}} \CM_i\geq \sum_{i=1}^p\frac{\pi^2}{\lambda^2[\Lambda_i]}.
\end{equation}


Notice, that we have some freedom in choosing the parametrization in \eqref{multi2}. Indeed, all the steps remain valid after
an application of any orthogonal rotation in the parameter space $\bvar'=O^{-1}\bvar$, $\bG'=O^T\bG$, which does not change the local cost function. Note, however, that the restriction to orthogonal transformations $O^TO=\openone$ is crucial here, as for a more general one $A$  the off-diagonal elements may appear in the formula for the cost $\tr(A^TA\cov)$, which would make it impossible to bound it using \eqref{mbounda}. Therefore, we can tighten the bound resulting from \eqref{mmmm} and write:
\begin{equation}
\label{mmbo}
\jntmm\gtrsim\max _O\frac{1}{N^2}\sum_{i=1}^p\frac{\pi^2}{\lambda^2([O^T\bG]_i)}.
\end{equation}

An analogous bound may be derived for the trace of the inverse of the QFI (see also \cite{Gorecki2020}):
\begin{equation}
\tr(F^{-1})=\sum_{i=1}^p [F^{-1}]_{ii}\geq \sum_{i=1}^p[F_i]^{-1}_{ii}\geq \sum_{i=1}^p[F_i^{-1}]_{ii}=\sum_{i=1}^p\frac{1}{n^2\lambda_i^2}
\end{equation}
so
\begin{equation}
\tr(F^{-1})\geq \max_O \frac{1}{n^2}\sum_{i=1}^p\frac{1}{\lambda^2([O^T\bG]_i)}
\end{equation}
and:
\begin{equation}
\label{crbo}
\jntcr\gtrsim \max_O \frac{1}{kn^2}\sum_{i=1}^p\frac{1}{\lambda^2([O^T\bG]_i)},
\end{equation}
where the same $O$ maximizes both \eqref{mmbo} and \eqref{crbo}.

\subsection{Relation between optimal global- and local- minimax costs}
\label{sec:locglob}
It is worth mentioning, that in the cases, where the local estimation problem may be extended to the covariant group estimation problem, the local minimax cost is the same as the one for estimating a completely unknown element of the group. Let us formalize it.

Let $U_g$ (where $g\in G$) be a unitary representation of compact group $G$ (so $U_{g_1}U_{g_2}=U_{g_1g_2}$) and consider the cost function invariant with respect to the action of this group $\forall_{g,\tilde g,h\in G}\cf(hg,h\tilde g)=\cf(g,\tilde g)$. Let $\bvar=[\var_1,...,\var_p]\mapsto g_{\bvar}$ be a local aprametrization around neutral element of the group $e\in G$ such that $U_{g_{\bvar}}=e^{i\bvar\bG}$. Then assuming that $\cf(e,g_{\tilde\bvar})=\tilde\bvar^2+o(\tilde\var_i\tilde\var_j)$, the asymptotic minimax cost for $\cf(g,\tilde g)$ for $g\in G$ is the same as the local one \eqref{mmlocal}; see \appref{App:cost} for more details. The reasoning is based on the same idea as the one performed for single parameter case~\cite{hayashi2011} (reminded here in section \ref{hlsp}).

Moreover, in~\cite{chiribella2008memory} it was shown that for covariant estimation the optimal results may be obtained within a parallel scheme (without the necessity of involving adaptiveness), which implies that also in the local minimax approach there is no advantage in applying adaptive strategy (in contrast to the results obtainable within CR formalism, which will be discussed in \ref{sec:commuting}).

\subsection{Separate strategy lower bound}
Finally, we would like to derive a simple bound for the minimal cost obtainable by the SEP+ strategy, which will allow us for a quick assessment of potential benefits due to a rotation in the parameter space. We have:
\begin{multline}
\seppmm\gtrsim \min_{A,N_j}\sum_{j=1}^p\frac{\pi^2}{N_j^2}\frac{[A^TA]_{jj}}{\lambda^2([A^T\bG]_j)}\\
\geq
\min_{A,N_j}\sum_{j=1}^p\frac{\pi^2}{N_j^2}\left(\min_i\frac{[A^TA]_{ii}}{\lambda^2([A^T\bG]_i)}\right)
\\
= \frac{p^3\pi^2}{N^2}\min_{A,i} \frac{[A^{T}A]_{ii}}{\lambda^2([A^T\bG]_i)}.
\end{multline}
Note, that the last term in above inequality depends only on the $i^{th}$ column of $A$.  Therefore, minimization over both $A$ and $i$ is equivalent to minimization over a single vector $\boldsymbol{a}$:
\begin{multline}
\min_{A,i}\frac{[A^{T}A]_{ii}}{\lambda^2([A^T\bG]_i)}=\min_{\boldsymbol{a}}\frac{|\boldsymbol{a}|^2}{\lambda^2[\boldsymbol{a}\bG]}\\=\min_{\boldsymbol{a}:|\boldsymbol{a}|^2=1}\frac{1}{\lambda^2[\boldsymbol{a}\bG]}=\frac{1}{\max\limits_{\boldsymbol{a}:|\boldsymbol{a}|^2=1}\lambda^2[\boldsymbol{a}\bG]},
\end{multline}
hence
\begin{equation}
\label{usemm}
\seppmm\gtrsim \frac{p^3\pi^2}{N^2}\frac{1}{\max\limits_{\boldsymbol{a}:|\boldsymbol{a}|^2=1}\lambda^2[\boldsymbol{a}\bG]}.
\end{equation}
Similarly, in the multiple repetition scenario we could write  (see also \cite{Gorecki2020}):
\begin{equation}
\label{usecr}
\seppcr\gtrsim \frac{p^2}{kn^2}\frac{1}{\max\limits_{\boldsymbol{a}:|\boldsymbol{a}|^2=1}\lambda^2[\boldsymbol{a}\bG]}.
\end{equation}

Thanks to the bounds derived in this section, we will be able to get an insight into the benefits of joint vs. separate strategies, even if we will not always be able to obtain a rigorous solution for the optimal achievable cost---see the next section.

\section{Examples}
\label{examples}
\begin{figure}[t]
\includegraphics[width=0.45 \textwidth]{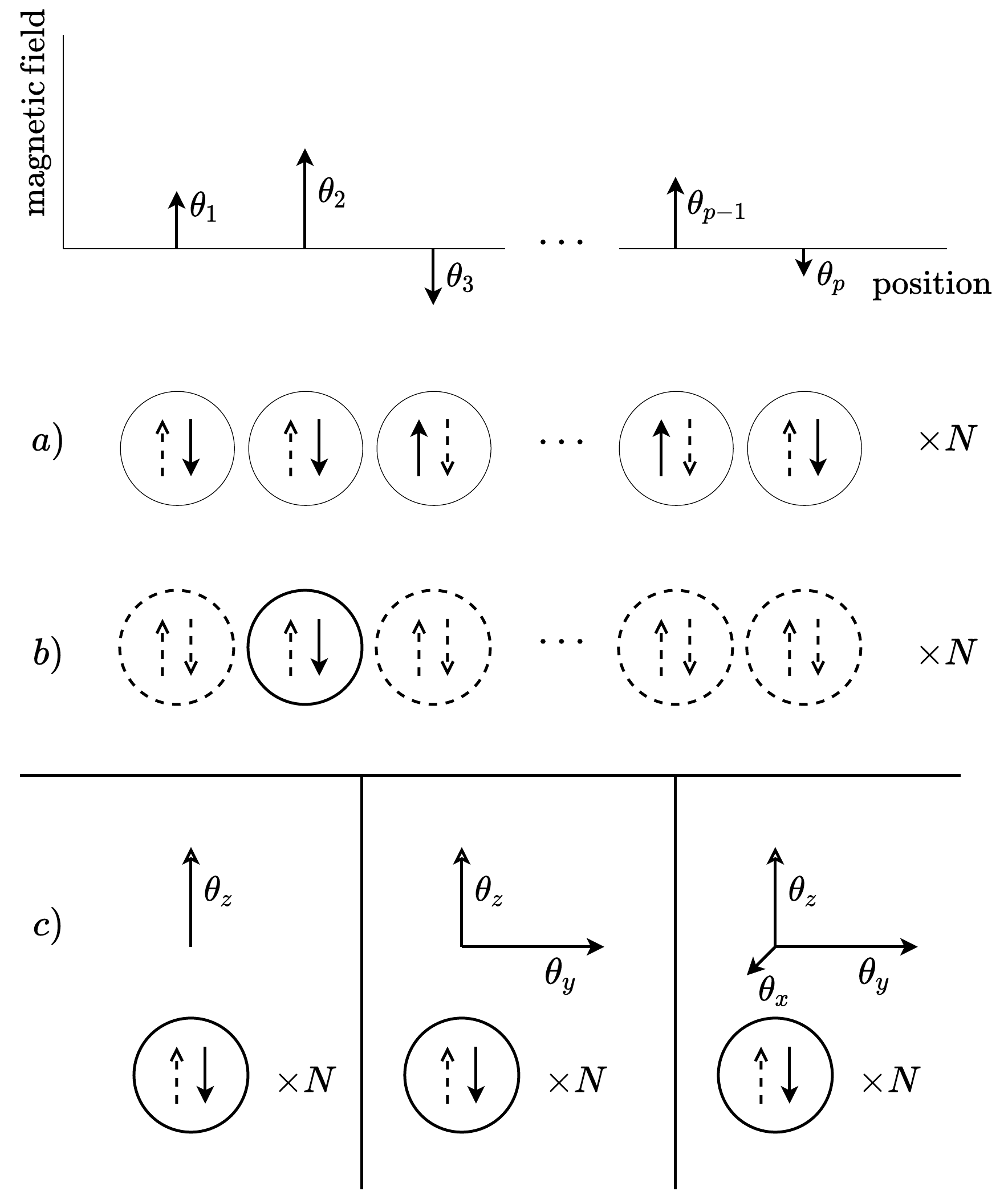}
\caption{Various models of magnetic field sensing by spin-$1/2$ atoms are discussed. In $a)$ and $b)$ the spatial distribution of magnetic field oriented in the $z$ direction is to be estimated. In  $a)$ the atoms are uniformly distributed in $p$ points and one has the freedom to choose their spin orientations in the optimal way, while in $b)$ for every single atom both the position and the spin orientation may be chosen arbitrary. In $c)$ the magnetic field in single point is measured, but multi-component estimation is discussed.}
\label{fig:systems}
\end{figure}

We will focus here on models which are inspired by various magnetic field sensing problems, but which are representative for
a wide range of multiparameter unitary estimation problems, see \figref{fig:systems}. As shown in \cite{gorecki2021multiple},
if all the generators mutually commute $\forall_i[\oG_i,\oG_j]=0$, there is no asymptotic advantage (for large $N$) in using a general adaptive strategy when compared to the parallel one. Therefore when analyzing the first two examples, we will focus on the parallel scheme only. For the last one, where the generators do not commute, both strategies will be discussed.

\subsection{Spatially distributed magnetic field sensing---fixed atoms positions}
\label{sec:fixed}
Consider the problem of spatially distributed magnetic field sensing (which is directed along the $z$ axis). The field is sensed by spin-$1/2$ atoms allocated in $p$ spatially separated places. 

Before moving on to the general solution of such a problem, we would like first to discuss it with an additional constraint imposed. Namely, we assume that the spatial distribution of sensing atoms is uniform and fixed, i.e. in each of $p$ points there are exactly $N$ atoms and one has only the freedom in choosing their spins orientations (this form of the problem was discussed in the many repetition scenario in~\cite{Gorecki2020}). Here, by an elementary amount of resources we understand a single layer of $p$-atoms, and the corresponding Hilbert space is the $2^p$-dimensional one spanned by the vectors of the form:
\begin{equation}
\ket{\boldsymbol{s}}=\ket{s_1,s_2,...,s_p},\quad \t{with}\quad s_i\in\{-1,+1\}.
\end{equation}
Then the elementary quantum gate is:
\begin{equation}
\label{fixed}
U_\bvar=e^{i\bvar\cdot\bG},\quad\Lambda_i= \openone^{\otimes i-1}\otimes \tfrac{1}{2}\sigma_z \otimes \openone^{\otimes p-i},
\end{equation}
so $\oG_i\ket{\boldsymbol{s}}=\frac{1}{2}s_i\ket{\boldsymbol{s}}$. Note that here the number of quantum gates $N$ is equal to number of $p$-atoms layers and in fact correspond to usage of $N\cdot p$ atoms.

Let us consider first the SEP strategy (without optimization over reparametrizations). Since $\forall_i \lambda[\oG_i]=1$,
one gets respectively: for the CR approach (using the analogues of $\noon$ \eqref{noon} states):
\begin{equation}
\sepcr\limeq p\times \frac{1}{k/p}\times\frac{1}{n^2}=\frac{p^2}{kn^2},
\end{equation}
and for the MM approach (with the use of $\siN$ \eqref{sin} states):
\begin{equation}
\sepmm\limeq p\times\frac{\pi^2}{(N/p)^2}=\frac{\pi^2p^3}{N^2}.
\end{equation}
However, looking at  \eqref{usemm} and \eqref{usecr} one may see that there is a significant potential for improvement in SEP+, as the value of $\lambda[\boldsymbol a \bG]$ will be maximal for $\boldsymbol a=1/\sqrt{p}\cdot[1,1,...,1]$ and equal to $\sqrt{p}$. Below we show a concrete reparametrization for which the mentioned bounds may be saturated. For simplicity, let us restrict to the case $p=2^r$ for some natural $r$.


In order to optimize the separate strategy one needs to find a parametrization for which each parameter may be sensed by all the atoms simultaneously. Therefore, instead of measuring the magnetic field point by point in $p$ positions, one may decompose the field into a proper components by the Walsh–Hadamard transformation $\bvar'=O^{-1}\bvar$ with:
\begin{equation}
\label{otrans}
O_{ij}=\frac{1}{\sqrt{p}}\prod_{k=0}^{r-1}(-1)^{i_kj_k}, \t{ where }\quad i=\sum_{k=0}^{r-1}i_k2^k,
\end{equation}
or, equivalently:
\begin{equation}
O=\frac{1}{\sqrt{p}}\begin{bmatrix}
1 & 1 \\
1 & -1
\end{bmatrix}^{\otimes r}.
\end{equation}
After the application of such a transformation, all the generators $[O^T\bG]_i$ remain diagonal in the basis $\{\ket{\boldsymbol{s}}\}$ and moreover:
\begin{equation}
\bra{\boldsymbol{s}}[O^T\bG]_i\ket{\boldsymbol{s}}=\frac{1}{2\sqrt{p}}\sum_{j=1}^p O_{ij}s_j.
\end{equation}
From that, indeed, $\forall_i \lambda([O^T\bG]_i)=\sqrt{p}$ and, moreover, as the eigenvectors of $\oG_i$ with minimal and maximal eigenvalues take the form:
\begin{equation}
\begin{split}
\ket{\lambda_{i+}}&=\ket{+O_{i1},+O_{i2},...,+O_{ip}}\\
\ket{\lambda_{i-}}&=\ket{-O_{i1},-O_{i2},...,-O_{ip}}
\end{split}
\end{equation}
the remaining generators acts on them trivially:
\begin{equation}
\oG_j\ket{\lambda_{i\pm}}=\pm\delta_{ij}\tfrac{1}{2}\ket{\lambda_{i\pm}}.
\end{equation}
Hence, when focusing on the estimation of a given parameter $\var_i$,
there are no disturbance issues related with the presence of the other parameters. Therefore:
\begin{equation}
\label{notfulsep}
\seppcr\limeq \frac{p}{kn^2},\quad \seppmm\limeq \frac{p^2\pi^2}{N^2}.
\end{equation}
We see that, thanks to the application of a proper reparametrization, we have decreased the cost obtainable in a separate strategy by a factor of $p$.
Note, that in order to use \eqref{otrans} we have assumed $p=2^r$. However, if this is not satisfied, one can still obtain qualitatively similar results, e.g. applying parameters transformation from \cite{Rey_1997}.

Going back to the initial parametrization, let us now discuss a joint strategy. As each of the parameters is associated
with a different atom, all of them may be measured without disturbing the measurement outcomes of the remaining ones. More formally, as the Hilbert space corresponding to the single layer of atoms has the characteristic structure $\mathcal H=(\mathbb C^2)^{\otimes p}$, then, for $N$ layers it may be written in the form $\mathcal H^{\otimes N}=\left((\mathbb C^2)^{\otimes p}\right)^{\otimes N}=\left((\mathbb C^2)^{\otimes N}\right)^{\otimes p}$. Hence, we may use the $\noon^{\otimes p}$ state (in the CR approach) or $\siN^{\otimes p}$ state (in the MM approach), which yields:
\begin{equation}
\label{notfuljnt}
\jntcr\limeq \frac{p}{kn^2},\quad \jntmm\limeq \frac{p\pi^2}{N^2},
\end{equation}
and which saturates the bound \eqref{mmbo} (which is the tightest for the original parametrization, i.e. with $O=\openone$).

Comparing \eqref{notfulsep} with \eqref{notfuljnt}, we see that for this model the existence of the advantage of measuring parameters jointly depends on the chosen paradigm---in the many repetition scenario there is no advantage, while for the fully optimal usage of all resources the advantage increases linearly with the number of parameters.


\subsection{Spatially distributed magnetic field sensing---arbitrary spatial distribution of atoms}
\label{sec:arbitrary}
Let us now consider the same problem with the full freedom in the distribution of the atoms in both space ($p$ positions) and spin orientations. The single atom Hilbert space will, therefore, be spanned by:
\begin{equation}
\ket{i,s},\quad \t{with}\quad i\in\{1,2,...,p\},\, s\in\{-1,+1\}.
\end{equation}
The corresponding single quantum gate has the form:
\begin{equation}
\label{free}
U_\bvar=e^{i\bvar\cdot\bG},\quad\oG_i=\ket{i}\bra{i}\otimes\tfrac{1}{2}\sigma_z=\tfrac{1}{2}(\ket{i,+}\bra{i,+}-\ket{i,-}\bra{i,-}),
\end{equation}
so $\oG_i\ket{j,s}=\delta_{ij}\frac{s}{2}\ket{j,s}$. Note that here, unlike in the previous example, the number of quantum gates is equal exactly to the number of atoms used.

Similarly as in \eqref{sin}, without loss we may restrict to the fully symmetric space, so that any $N$-atomic output state may be written as:
\begin{equation}
\label{sst}
\ket{\psi^{N}_{\bvar}}=\sum_{|\boldsymbol m|=N} e^{\tfrac{i}{2}\sum \var_i(m_{i+}-m_{i-})}c_{\boldsymbol m}\ket{\boldsymbol m},
\end{equation}
where $\boldsymbol m=[m_{1+},m_{1-},m_{2+},...,m_{p-}]$ and $|\boldsymbol m|=\sum_{i=1}^p(m_{i+}+m_{i-})$.

In this case $\forall_i \lambda[\oG]=1$---their non-zero eigenspaces are mutually orthogonal and hence cannot be increased by taking any linear combination $\boldsymbol{a}\bG$. Therefore, there in no advantage of the SEP+ over the SEP protocol, and we simply get:
\begin{equation}
\sepcr\limeq \frac{p^2}{kn^2}
\end{equation}
for the state:
\begin{equation}
\label{1dnoon}
\ket{\psi^{n}_{\bvar}}=\frac{1}{\sqrt{2}}\left(e^{in\var_i/2}\ket{n}_{i+}\ket{0}_{i-}+e^{-in\var_i/2}\ket{0}_{i+}\ket{n}_{i-}\right),
\end{equation}
while in case of full optimization paradigm we get
\begin{equation}
\sepmm\limeq \frac{p^3\pi^2}{N^2}
\end{equation}
for the state:
\begin{multline}
\label{1dopt}
\ket{\psi^{N/p}_{\bvar}}=\sum_{m=0}^{N/p}\frac{\sqrt{2}e^{i(2m-N/p)/2\var_i}}{\sqrt{N/p+2}}\\
\sin\left(\frac{(m+1)\pi}{N/p+2}\right)\ket{m}_{i+}\ket{N/p-m}_{i-},
\end{multline}
where $\ket{m}_{i+}\ket{N/p-m}_{i-}$ denotes a state with  $N/p$ atoms in the $i$-th position, with $m$ of them being oriented up, and $N/p-m$ down.

Much more interesting aspects may be observed when analyzing the joint strategy. First, let us look what insight may be obtained from the bounds \eqref{crbo} and \eqref{mmbo}. To do so, we will use again the Walsh-Hadamard transformation. However, note that in this case the effect is completely opposite to the one observed in the  previous example---application of this transformation \textbf{decreases} all $\lambda[O^T\oG_i]=1/\sqrt{p}$---that is, in the case where one has the full freedom of distributing atoms in space,
measuring combinations of magnetic fields at various points, instead of measuring the field point by point, is an inefficient separate strategy. Still, thanks to this obervation, such a transformation may be used to tighten the bounds \eqref{crbo}, \eqref{mmbo}.
Since $\sum_{i=1}^p 1/\lambda^2([O^T\bG]_i)=p^2$, it gives:
\begin{equation}
\label{frebo}
\jntcr\gtrsim \frac{p^2}{kn^2},\quad \jntmm\gtrsim \frac{p^2}{N^2}.
\end{equation}
In the CR case, the above bound may be saturated using the state:
\begin{equation}
\ket{\psi^n_\bvar}=\frac{1}{\sqrt{p}}\sum_{i=1}^p\tfrac{1}{\sqrt{2}}\left(e^{+in\var_i/2}\ket{n}_{i,+}+e^{-in\var_i/2}\ket{n}_{i,-}\right),
\end{equation}
which is simply an equally weighted superposition of \eqref{1dnoon} and for which:
\begin{equation}
F_{ij}=4\t{Re}\left[\braket{\psi_i|\psi_j}-\braket{\psi_i|\psi}\braket{\psi|\psi_j}\right]\Rightarrow F_{ij}=\frac{\delta_{ij}}{p},
\end{equation}
and consequently
\begin{equation}
\tr(F^{-1})=\frac{p^2}{n^2}\Rightarrow \jntcr\limeq \frac{p^2}{kn^2}.
\end{equation}

To analyze the joint strategy within the MM approach, we will use the Fourier analysis, originally applied to the single parameter estimation problem in \cite{Imai2009}, later generalized for another group estimation problems \cite{hayashi2016fourier}, and very recently applied for multiphase estimation in multiarm interferometer with constraining for the total number of photons~\cite{gorecki2021multiple}. Note that from the point of view of the discussed problem, only the differences between the number of atoms oriented up and down $\Delta m_i=(m_{i+}-m_{i-})/2$ matter. Therefore, for a given state \eqref{sst} we define normalized states $\ket{\Delta \boldsymbol m}$ and coefficients $c_{\Delta \boldsymbol m}$ satisfying:
\begin{equation}
\forall_{\Delta \boldsymbol m} c_{\Delta \boldsymbol m}\ket{\Delta \boldsymbol m}=\sum_{\boldsymbol m: \forall (m_{i+}-m_{i-})/2=\Delta m_i} c_{\boldsymbol m}\ket{\boldsymbol m}.
\end{equation}
Then \eqref{sst} may be rewritten in the form:
\begin{equation}
\ket{\psi_\bvar^N}=\sum_{\sum_i |\Delta m_i|\leq N/2} e^{i\bvar \Delta \boldsymbol m }
c_{\Delta \boldsymbol m}\ket{\Delta \boldsymbol m}.
\end{equation}

Next, for $N$ large enough, we may replace discrete variables by continuous ones $\frac{\Delta m_i}{N}\to\mu_i\in[-1/2,+1/2]$ to get:
\begin{equation}
\label{statecont}
\ket{\psi_{f,\bvar}^N}=\int\limits_{\sum_i|\mu_i|\leq 1/2} d{\boldsymbol \mu} e^{iN\bvar{\boldsymbol \mu}}f({\boldsymbol \mu})\ket{{\boldsymbol \mu}}.
\end{equation}
As argued in \cite{gorecki2021multiple}, the optimal measurement in the asymptotic limit will be the covariant one:
\begin{equation}
\ket{\chi_{\tilde\bvar}}=\frac{1}{\sqrt{(2\pi/N)^p}}\int \t{d}\boldsymbol{\mu}\, e^{iN\boldsymbol{\mu}\tilde\bvar}\ket{\boldsymbol{\mu}}.
\end{equation}
For technical reasons, in further calculations we will treat the function $f({\boldsymbol \mu})$ appearing in \eqref{statecont} as the one defined on the whole $\mathbb R^p$, but equal zero everywhere outside of $\{[\mu_1,..,\mu_p]\}_{\sum_i|\mu_i|\leq 1/2}$ (which allows us to perform the standard Fourier transform of this function). The mean value of the quadratic cost is given by:
\begin{equation}
\int\displaylimits_{\mathbb R^p} \t{d} \tilde{\bvar}\, |\braket{\chi_{\tilde{\bvar}}|{\psi_{f,\bvar}^N}}|^2 (\tilde\bvar-\bvar)^2  = \frac{1}{N^2}\int\displaylimits_{\mathbb R^p} \t{d} \tilde{\bvar}\,
\left|\hat{f}(\tilde\bvar)\right|^2\tilde\bvar^2,
\end{equation}
where $\hat{f}$ is the Fourier transform of $f$ and we dropped the irrelevant dependence on $\bvar$.

Going back to the $\boldsymbol{\mu}$-representation and performing the minimization over $f$ we get:
\begin{equation}
\label{tominimize}
\begin{split}
\jntmm\limeq &\frac{1}{N^2}\min_{f}\int\limits_{\sum_i|\mu_i|\leq 1/2} \t{d}\boldsymbol{\mu}\, f^*(\boldsymbol{\mu})\left(\sum_{k=1}^p-\partial_{\mu_k}^2\right)f(\boldsymbol{\mu}),\\
&{\rm with}\quad \int\limits_{\sum_i|\mu_i|\leq 1/2} \t{d}\boldsymbol{\mu}\, |f(\boldsymbol{\mu})|^2=1,\\
&f(\boldsymbol{\mu})=0\quad \textrm{for } \boldsymbol{\mu} \textrm{ on the boundary } \sum_i |\mu_i|=1/2.
\end{split}
\end{equation}

\begin{figure}[t]
\includegraphics[width=0.45 \textwidth]{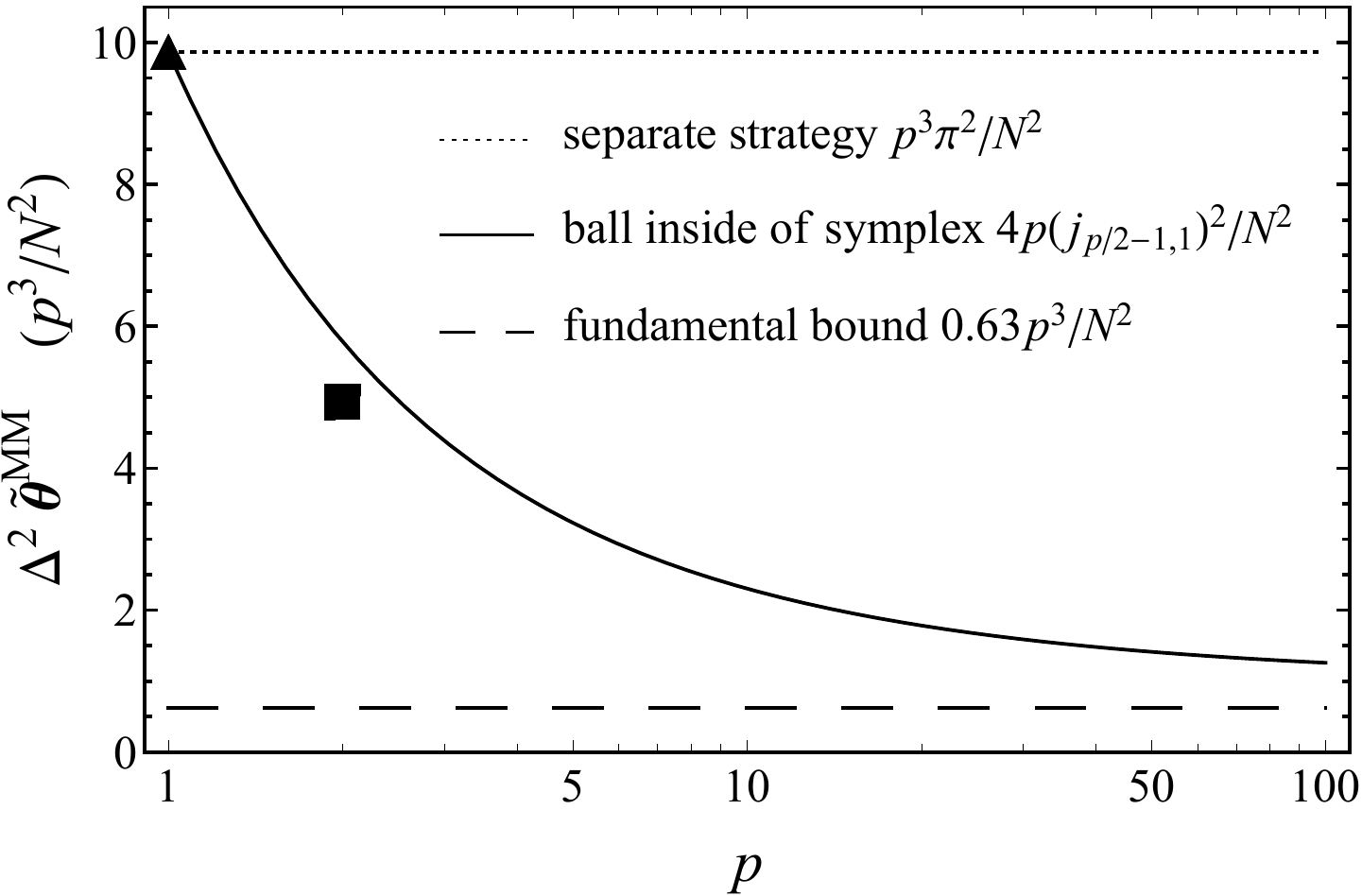}
\caption{The minimal achievable cost in the problem of estimating spatial distribution of magnetic field in $p$ point is analyzed in the limit of large $N$ (with no repetition scenario). The triangle and square represent analytically found results for $p=1,2$. The solid line represent the variance obtainable by exemplary suboptimal state state. The dotted line corresponds to the separate strategy, while the dashed one is the fundamental (not necessary obtainable) bound \eqref{anotherbound}.}
\label{fig:ball}
\end{figure}

The problem is therefore equivalent to minimization of the kinetic energy of a particle in infinite potential well in a shape of a $p$-dimensional simplex. The analytical solutions are known only for $p=1,2$ (see \appref{App:airy}). For higher number of parameters we will derive a lower bound. As $f(\boldsymbol{\mu})=0$ everywhere outside of $\sum_i|\mu_i|\leq 1/2$,  the mean value of $\sum_i|\mu_i|$ is trivially smaller or equal $1/2$. Next, thanks to the symmetry, we may assume without loss for generality, that the function $f(\mu)$ minimizing the above is fully symmetric under the exchange of variables $\mu_i$ and therefore all the mean values of $|\mu_i|$ are equal and $\leq 1/(2p)$
(see also supplementary materials of \cite{gorecki2021multiple} for a broader discussion of this argument).
The minimal sum of variances may be therefore be bounded as:
\begin{equation}
\label{singleminimze}
\jntmm\gtrsim p\times \min_g\int_{-\infty}^{+\infty}d\mu g^*(\mu)\left(-\frac{\partial^2}{\partial^2\mu}\right)g(\mu)
\end{equation}
with
\begin{equation}
\int_{-\infty}^{+\infty}d\mu |g(\mu)|^2=1,\quad \int_{-\infty}^{+\infty}d\mu |g(\mu)|^2|\mu|=\frac{1}{2p}.
\end{equation}
Moreover, from the symmetry of the problem, the solution will be symmetric with respect to the $0$ point (which, assuming differentiability, implies $\frac{\partial g}{\partial\mu}\Big|_{\mu=0}=0$). Therefore, the problem is equivalent to:
\begin{equation}
\label{singmin}
\begin{split}
&\min_g\int_{0}^{+\infty}d\mu g^*(\mu)\left(-\frac{\partial^2}{\partial^2\mu}\right)g(\mu),\quad \t{with}\quad \frac{\partial g}{\partial\mu}\Big|_{\mu=0}=0,\\
&\int_{0}^{+\infty}d\mu |g(\mu)|^2=1,\quad \int_{0}^{+\infty}d\mu |g(\mu)|^2|\mu|=\frac{1}{2p},
\end{split}
\end{equation}
which may be solved using the Lagrange multiplayers method. The solution in terms of the Airy function yields the final bound (see \appref{App:airy} for detail derivation):
\begin{equation}
\label{anotherbound}
\jntmm\gtrsim \frac{0.63p^3}{N^2}.
\end{equation}
We are unable to proof the tightness of the above bound. However, we are able to point our an  exemplary state for which the cost closely approaches the bound. Consider the largest possible $p$-dimensional ball inside the simplex $\sum_i|\mu_i|\leq 1/2$ and then as  $f(\boldsymbol \mu)$ choose the function which minimizes the kinetic energy inside this ball with the boundary condition $f(\boldsymbol \mu)=0$ on the border and outside of the ball. The cost corresponding to this construction, the bound \eqref{anotherbound}, the values of analitical solution of \eqref{tominimize} for $p=1,2$ as well as $\sepmm$ are plotted together in \figref{fig:ball}---details of the calculations may be found in \appref{App:airy}. For large $p$, the total cost corresponding to the described strategy leads to $p^3/N^2$, and we finally get:
\begin{equation}
\jntmm\limeq \frac{cp^3}{N^2},\quad 0.63\leq c_1\leq 1.
\end{equation}
When comparing to \eqref{frebo}, we see that in the MM scenario not only the bound is not tight, by it even fails to properly predict the scaling of the cost with the number of parameter $p$.

To summarize, for the problem of estimation of the spatially distributed single component magnetic filed, with full freedom in choosing both the position and orientation of the atoms, for large $p$ we obtain:
\begin{equation}
\label{uno}
\begin{split}
\sepcr\limeq \frac{p^2}{kn^2},\quad & \sepmm\limeq \frac{\pi^2p^3}{N^2}\\
\jntcr\limeq \frac{p^2}{kn^2},\quad & \jntmm\limeq \frac{c_1p^3}{N^2}\\
& \quad 0.63\leq c_1\leq 1.
\end{split}
\end{equation}

If one wants to compare these results with the previous example (where the positions of the atoms were fixed), one should bare in mind, that in the previous example one gate corresponded to $p$ atoms, not one. Therefore, in order to make the comparison fair, one should rewrite \eqref{notfulsep}, \eqref{notfuljnt} in terms of the atoms used $N_a=p\cdot N$, $n_a=p\cdot n$, which gives:
\begin{equation}
\label{dos}
\begin{split}
\seppcr\limeq \frac{p^3}{kn_a^2},\quad & \seppmm\limeq \frac{\pi^2p^4}{N_a^2}\\
\jntcr\limeq \frac{p^3}{kn_a^2},\quad & \jntmm\limeq \frac{\pi^2p^3}{N_a^2}.
\end{split}
\end{equation}
Then it is clear, that for the same amount of atoms used, all the costs from \eqref{dos} are larger than the corresponding ones from \eqref{uno}, as in \eqref{dos} fewer degrees of freedom are allowed.


\begin{figure}[t]
\includegraphics[width=0.45 \textwidth]{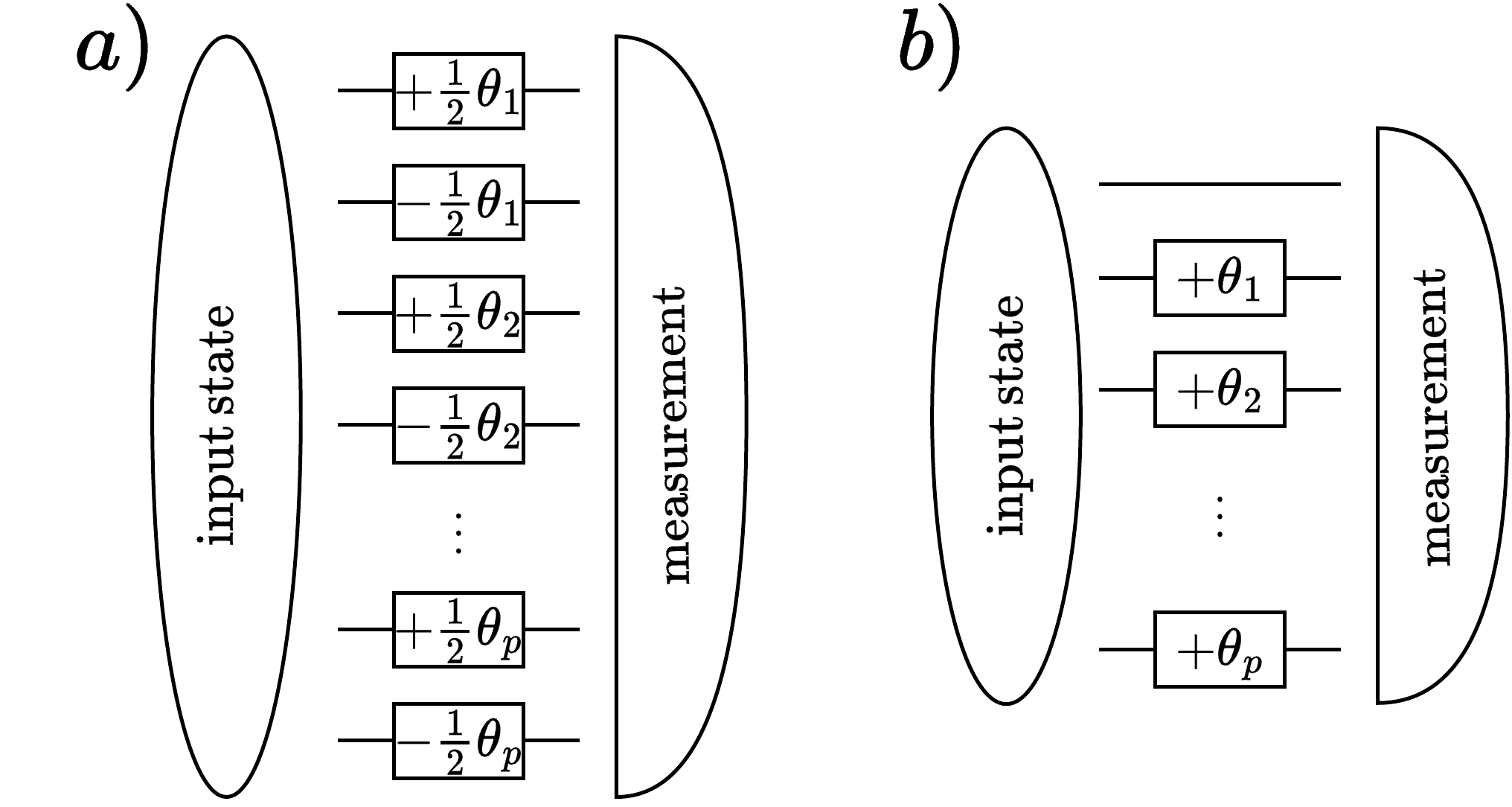}
\caption{The problem of estimating spatial distribution of magnetic field sensed by $N$ spin-$1/2$ atoms, with full freedom in choosing their position and orientation may be equivalently seen as the problem of $p$ phases sensing in a $2p$ arms interferometer using $N$ photons (diagram $a$). This is a slightly modificated version of the problem discussed in~\cite{gorecki2021multiple}, where $p$ phases where measured in the presence of a single reference arm in $(p+1)$-arms inteferometer (diagram $b$).
}
\label{fig:phases}
\end{figure}

Finally, one may notice that the problem of estimating magnetic field in $p$ points by $N$ atoms (with full freedom in choosing their spin and position) is equivalent to a slightly modified problem of multi-phase estimation discussed in \cite{gorecki2021multiple}. Indeed, treating the points in space as arms of an interferometer and atoms as photons, we may think about this problem in terms of a $2p$-arm inferferometer with   phase shifts $\pm \frac{1}{2}\var_i$ in each arm. Therefore, it makes sense to compare \eqref{uno} with the analogue costs obtained in \cite{gorecki2021multiple} for the problem of estimating $p$ unknown phase-shifts in a $(p+1)$-arm interferometer (where the one arm is the reference arm)---both version are schematically presented in \figref{fig:phases}. In the latter case, the single photon Hilbert space is spanned by $\{ \ket{0},\ket{1},..,\ket{p}\}$ (where $\ket{0}$ corresponds to the reference arm, and $\ket{i}$ to the sensing arms), and the quantum gate is $U_\bvar=\exp(i\sum_{i=1}^p \var_i \ket{i}\bra{i})$. For large $p$:
\begin{equation}
\label{tres}
\begin{split}
\sepcr\limeq \frac{p^2}{kn^2},\quad & \sepmm\limeq \frac{\pi^2p^3}{N^2}\\
\jntcr\limeq \frac{p^2}{4kn^2},\quad & \jntmm\limeq \frac{c_2p^3}{N^2}\\
& \quad 1.89\leq c_2\leq 2.
\end{split}
\end{equation}
Comparing \eqref{uno} with \eqref{tres}, we may notice significant differences. While in the $(p+1)$-arm interferometer, the problem in both the CR and the MM approach, reveals a constant advantage of JNT over SEP strategies of a similar order (around $5$), for \eqref{uno} no advantage is observed in the CR approach, while the one in the MM approach is even larger (around $10$).

This effect, however, may be easily understood after analyzing the structure of the states optimal for measuring parameter separately---\eqref{1dnoon} and \eqref{1dopt}.
While in the case studied in \cite{gorecki2021multiple}, in order to estimate any of the parameters one needed to use the same reference arm, in the presently studied model each phase may be measured using two dedicated arms with phase shifts $[-\var_i/2]$ and $[+\var_i/2]$. Consequently, the corresponding $\noon$ states \eqref{1dnoon} are mutually orthogonal---the resources ``consumed'' by one parameter cannot be used to estimate others.
On the other hand, looking at $\siN$ state \eqref{1dopt} we see that the greatest weights are attached to the vectors with relatively small differences $m_{i+}-m_{i-}=0$. As there is no problem in distributing photons (or atoms) in such a way that this difference is small for all $i$, such a component may be used in estimating all the parameters simultaneously, which is responsible for the significant advantage of the joint strategy in this case.

\subsection{Multicomponent magnetic field estimation}
\label{sec:commuting}
As the last problem let us discuss the canonical example of estimation of parameters associated with non-commuting generators.
More specifically, we focus on the problem of estimating the three components of a magnetic field vector $\bB=[\B_1,\B_2,\B_3]^T$ in a given point in space using spin-$1/2$ atoms. Single atom Hilbert space is therefore simply a qubit space and the corresponding quantum gate reads:
\begin{equation}
U_{\bB}=e^{i\bB\cdot\boldsymbol{\sigma}/2},
\end{equation}
where $\boldsymbol\sigma=[\sigma_1,\sigma_2,\sigma_3]^T$ is the vector of Pauli matrices. Unlike in the previously discussed examples, here the minimal achievable cost depends on the actual values of the parameters $\bvar$~\cite{baumgratz2016}. For simplicity, let us focus on estimation around point $\bB_0=[0,0,0]^T$.

Note that for any normalized vector $\boldsymbol{a}$, the operator $\boldsymbol{a}\cdot\boldsymbol{\sigma}$ has the same eigenvalues. Hence,
invoking from \eqref{usemm}, \eqref{usecr} we see that a rotation in the parameter space cannot improve the precision in a separate protocol---there is no advantage in SEP+.

Let us start by discussing the many repetition scenario. In a separate strategy, each component may be measured with a proper n00n states $\frac{1}{\sqrt{2}}(\ket{+}_{x,y,z}^{\otimes n}+\ket{-}_{x,y,z}^{\otimes n}$), which leads to:
\begin{equation}
\sepcr\limeq 3\times\frac{1}{k/3}\times\frac{1}{n^2}=\frac{9}{kn^2}.
\end{equation}

For the joint strategy in turns out that, unlike in the previously discussed examples, application of the adaptive scheme with ancillas allows to beat the performance of the optimal parallel strategy. We will, therefore, discuss these two strategies independently.


The minimal trace of inverse of the QFI achievable for parallel scheme may be found analytically~\cite{kolenderski2008} (for $n\geq 6$): $\tr(F_{\rm parallel}^{-1})=\frac{9}{n(n+2)}$ (and condition \eqref{imag} is satisfied), so
\begin{equation}
{\jntcr}_{\t{parallel}}\limeq \frac{9}{kn(n+2)}\overset{n\gg 1}{\approx}\frac{9}{kn^2},
\end{equation}
which for large $n$ is almost the same as for the separate strategy. Therefore, the advantage offered by joint measurement in the parallel strategy disappears with increasing $n$.

In contrast to the above, it was shown in~\cite{Yuan2016}, that for the adaptive ancilla-assisted sequential scheme utilizing as an input state
\begin{equation}
\ket{\psi_{\ins}}=\frac{1}{\sqrt{2}}(\ket{+}_z\ket{0}_A+\ket{-}_z\ket{1}_A)
\end{equation}
(where $\ket{0}_A,\ket{1}_A$ belongs to ancillary system) and acting on in by the gate $n$ times, one may obtain $\tr(F_{\rm adaptive}^{-1})=\frac{3}{n^2}$ (satisfying \eqref{imag}), and hence
\begin{equation}
{\jntcr}_{\t{adaptive}}\limeq \frac{3}{kn^2},
\end{equation}
which saturates \eqref{mmbo}. Therefore, we see that in the many repetition scenario, the possibility of acting sequentially is crucially needed to take the advantage from the joint estimation approach as it allows to decrease the final variance by a factor $3$, compared to the optimal separate or joint parallel strategy.

Consider now the fully optimal usage of $N$ gates. Note that, unlike the previously discussed examples, here the existence of unknown parameters $\var_{j\neq i}$ may significantly impede estimation of $\var_i$, as it may be impossible to find an initial state, for which the evolution would depend only on $\var_i$. As the consequence, we can only write the lower bound:
\begin{equation}
\sepmm\gtrsim 3\times \frac{\pi^2}{(N/3)^2}=\frac{27\pi^2}{N^2}.
\end{equation}
Still, it is not obvious how to obtain such a precision while measuring parameterss separately.

To calculate the asymptotically optimal cost in joint estimation we use the fact, that the problem may be extended to the covariant one, which allows us to use the reasoning from \ref{sec:locglob}, stating that optimal asymptotical local minimax cost is the same as the global one, obtainable within the parallel scheme.
In~\cite{chiribella2004,chiribella2005su2}, the estimation of a completely unknown element of $SU(2)$ within the parallel strategy was discussed with the covariant cost
\begin{equation}
\label{su3}
e(\bvar,\tilde\bvar)=6-2\tr(U^{(1)}_\bvar U^{(1)\dagger}_{\tilde\bvar})\overset{|\bvar|,|\tilde\bvar|\ll 1}{\approx}2|\bvar-\tilde\bvar|^2,
\end{equation}
where $U^{(1)}_\bvar$ is a rotation matrix of a spin-1 particle. It was shown that the asymptotic minimal cost is $e(\bvar,\tilde\bvar)\limeq 8\pi^2/N^2$, which is achievable using initial state:
\begin{equation}
\label{3dopt}
\ket{\psi_{\ins}}=\sqrt{\frac{2}{N/2+1}}\sum_{j=0(\tfrac{1}{2})}^{N/2-1}\sin\left(\tfrac{(j+1)\pi}{J+1}\right)
\left(\sum_{\alpha=1}^{2j+1}\frac{\ket{j\alpha,m_j=\alpha}}{\sqrt{2j+1}}\right),
\end{equation}
where $\ket{j\alpha,m_j=\alpha}$ are states with a well defined total angular momentum $j$ and its projection onto $z$ direction, while $\alpha$ numerates different subspaces corresponding to equivalent irreducible representations of $SU(2)$. Therefore, including the factor $2$ that appears in \eqref{su3} we get:
\begin{equation}
\label{su2res}
\jntmm\limeq \frac{4\pi^2}{N^2},
\end{equation}
which shows a significant advantage over the optimal separate strategy. Based on the discussion in \ref{sec:locglob} and ~\cite{chiribella2008memory}, this result, which is already obtainable within a parallel scheme, cannot be improved by applying any adaptive strategy. This is in stark contrast to many repetition scenario, where a parallel joint strategy offers no asymptotical advantage with respect to a separate one, while adaptiveness allows decreasing the cost by a factor of  $3$. It is also worth noting that the result \eqref{su2res} may be also obtained for parallel strategies with the usage of Fourier analysis, as shown in \cite[Section 12]{hayashi2016fourier}.

Analogous reasoning may be performed in a situation, when one component of the magnetic field is known to be zero, and only the two remaining components are being estimated.
For the separate strategy we get:
\begin{equation}
\sepcr\limeq \frac{4}{kn^2},\quad \sepmm\gtrsim\frac{8\pi^2}{N^2},
\end{equation}
for the joint CR:
\begin{equation}
{\jntcr}_{\t{parallel}}\limeq \frac{4}{kn(n+2)},\quad{\jntcr}_{\t{adaptive}}\limeq \frac{2}{kn^2},
\end{equation}
while for the joint MM:
\begin{equation}
\label{su2res2}
\jntmm\limeq \frac{4\xi^2}{N^2}\approx \frac{2.34\pi^2}{N^2},
\end{equation}
where the parallel strategy obtaining above was found in~\cite{bagan2000,bagan2001} and $\xi\approx 2.4048$ is first zero of the Bessel function $J_0(x)$. \eqref{su2res2}, similarly like \eqref{su2res}, is valid for both parallel and sequential adaptive strategies.

If, on the other hand, the direction of the magnetic field is known and only the length of the magnetic vector is to be estimated, the problem is equivalent to single phase estimation problem discussed before.


A natural extension of all the above considerations would be to combine all the examples, and consider the most general problem of
estimating all the 3 components of a spatially distributed magnetic field. Based on the analysis performed we expect no improvement in the CR approach and some constant improvement in the MM approach. The strict analysis of this problem, however, is beyond of scope of this paper.

\section{Conclusions}
\label{sec:con}
The examples studied, demonstrate that, unlike in the single unitary parameter case, in a multiparameter estimation problem there is no simple correspondence between the results obtained within the many repetition paradigms and the one, where all the resources are accumulated in single experimental realization. In the case,  where the total amount of resources is limited (no matter how large), the analysis based only on the QFI is not sufficient to draw not only quantitative but even qualitative conclusions. The presented examples showed, that such an approach tend to overrate the performance of SEP/SEP+ strategies and this opens up the possibility that certain joint estimation metrological strategies may offer a significant advantage, even if this is not apparent in a formalism based on the QFI.

\begin{acknowledgments}
We thank Howard M. Wiseman for many fruitful discussions.
This work was supported by the National Science Center (Poland) Grant No. 2020/37/B/ST2/02134. Wojciech G{\'{o}}recki was also supported by the Foundation for Polish Science (FNP) via the START scholarship.
\end{acknowledgments}

%

\appendix

\renewcommand{\thesection}{\Alph{section}}
\setcounter{theorem}{0}
\renewcommand{\thetheorem}{A\arabic{theorem}}

\onecolumngrid

\section{Different approach to the identification of the advantage of joint estimation protocols}
\label{App:incompatibility}
Alternatively to the approaches described in the main text, the issue of the potential gain comming from measuring multiple parameters simultaneously versus measuring them separately may be explored abstracting from the problem of optimal division of resources, but instead by analyzing the so called probe incompatibility \cite{ragy2016,albarelli2021probe}. In this approach the minimal cost achievable in a joint strategy is compared with the one coming from measuring each parameter individually, but with the assumption, that in the latter case for each parameter one spends the same amount of resources as in the whole join strategy (so effectively in the separate strategy $p$ times more resources are consumed).

While in the current paper the potential superiority of joint measurement is discussed for a particular cost function (defined by the chosen parametrization for which it is equal to identity), one may instead try to look at the problem as the feature of the channel itself. To do so, in~\cite{albarelli2021probe} the following quantity was introduced:
\begin{equation}
\label{franc}
\mathfrak J^*=\max_{\{\bold{w}_i\}}\left(\frac{\min\limits_{(n,k)-\t{protocol}}\tr(W\cov')}{\sum_i\min\limits_{(n,k)-\t{protocol}}\bold{w}_i^T\cov'\bold{w}_i}\right),
\end{equation}
where $W=\sum_i \bold{w}_i\bold{w}_i^T$. Note that when compared to the original notation~\cite{albarelli2021probe}, we have added a prime sign to $\cov'$ and $F'$, as we reserve  the un-primed parametrization for the case where the cost matrix is the identity.

In order to understand this approach, in the context of \figref{fig:scheme}, let us perform a reparametrization $\bvar=A\bvar'$ with the transformation matrix $A=[\bold{w}_1,...,\bold{w}_p]^T$. Then, for any fixed set $\{\bold{w}_i\}$ we have:
\begin{equation}
\frac{\min\limits_{(n,k)-\t{protocol}}\tr(W\cov')}{\sum_i\min\limits_{(n,k)-\t{protocol}}\bold{w}_i^T\cov'\bold{w}_i}=
\frac{\min\limits_{(n,k)-\t{protocol}}\tr(\cov)}{\sum_i\min\limits_{(n,k)-\t{protocol}}\cov_{ii}}=
\frac{\jntcr}{\sum_i\min\limits_{(n,k)-\t{protocol}}\Delta^2\tilde\var_i}
\end{equation}
so for a fixed $\{\bold{w}_i\}$ it corresponds to a situation, where Bob alternatively sends $n\cdot k$ gates to Alice (nominator) or the same amount of gates directly to each of the Minions (denominator). In this sense the denominator is similar to the SEP strategy (but with omitted problem or resources distribution). Note, however, that the maximization over $\{\bold{w}_i\}$ in \eqref{franc} \textbf{does not} correspond to SEP+ (as here also the value in the nominator changes while for the reparametrization of $A$ in SEP+ it does not)---it should be rather seen as maximization over all possible Bob's initial parametrizations.

\section{Necessity of a non-orthogonal transformation in order to obtain the minimal SEP+ cost---example}
\label{App:A}
\begin{figure}[t]
\includegraphics[width=0.45 \textwidth]{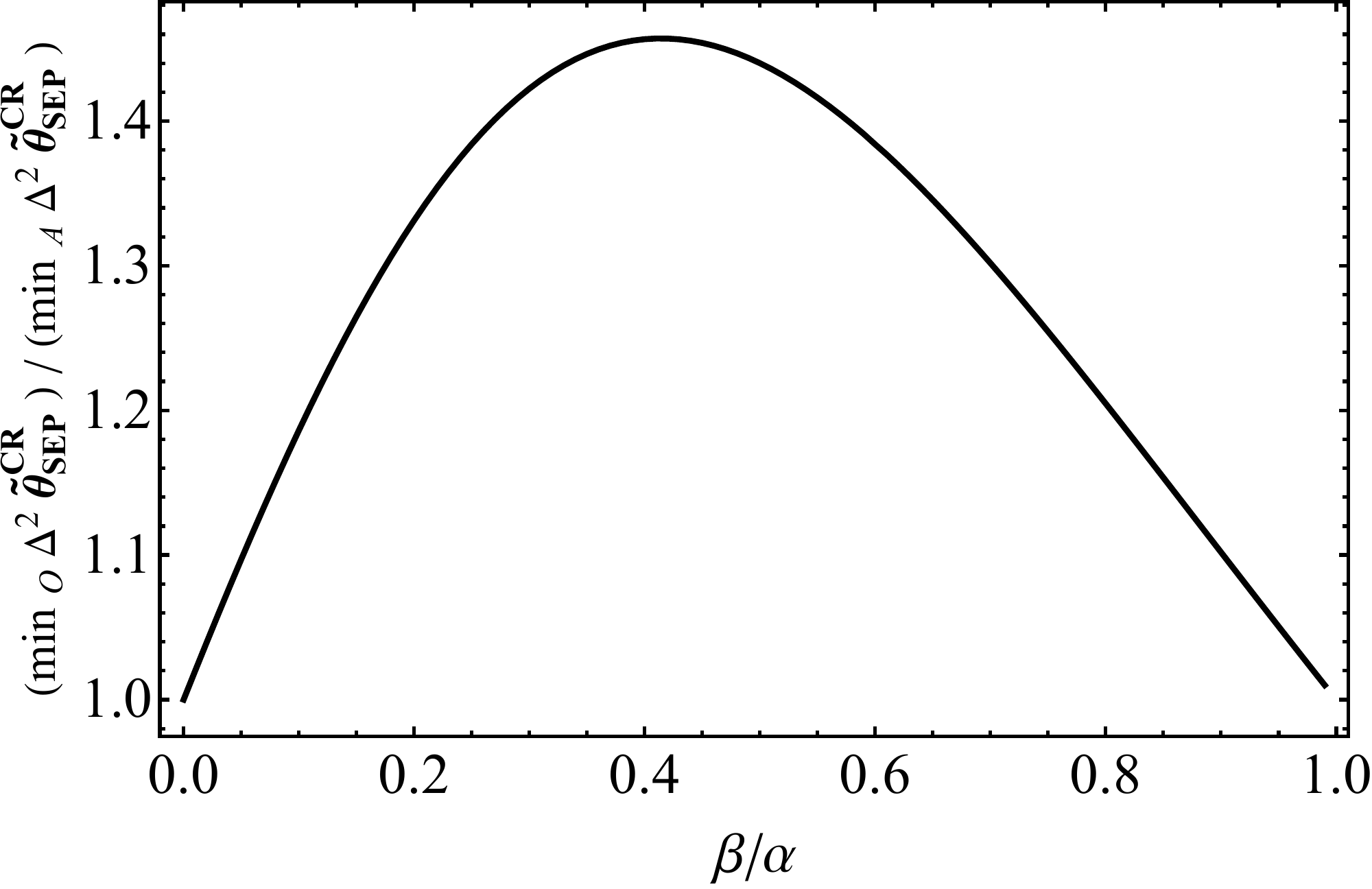}
\caption{The ratio between the minimal cost in a separate strategy achievable via optimization over orthogonal reparametrization and
a general one.}
\label{fig:A}
\end{figure}

Here we discuss an exemplary two-parameter estimation problem, for which in order to obtain the minimal cost in SEP+ protocol we need to apply a non-orthogonal transformation $A$ in the parameter space. We will focus on the many repetition paradigm and the CR formalism. Consider a unitary channel:
\begin{equation}
U_\bvar=e^{i\bvar\cdot\bG},\quad \oG_1=\tfrac{1}{2}\diag(+\alpha,-\alpha,+\beta,-\beta),\quad \oG_2=\tfrac{1}{2}\diag(+\beta,-\beta,+\alpha,-\alpha),
\end{equation}
where $0<\beta<\alpha$ and $\diag(...)$ is a diagonal matrix acting on the Hilbert space spanned by $\ket{1},\ket{2},\ket{3},\ket{4}$, so
\begin{equation}
\bvar\cdot\bG=
\begin{bmatrix}
+\tfrac{1}{2}(\alpha\var_1+\beta\var_2) & 0 & 0 & 0\\
0 & -\tfrac{1}{2}(\alpha\var_1+\beta\var_2) & 0 & 0\\
0 & 0 & +\tfrac{1}{2}(\beta\var_1+\alpha\var_2) & 0\\
0 & 0 & 0 & -\tfrac{1}{2}(\beta\var_1+\alpha\var_2)\end{bmatrix}.
\end{equation}
Since the generators mutually commute, the Fisher information matrix for the input $\ket{\psi}$ is given by:
\begin{equation}
[F]_{ij}=4\left(\bra{\psi}\oG_i\oG_j\ket{\psi}-\bra{\psi}\oG_i\ket{\psi}\bra{\psi}\oG_j\ket{\psi}\right)
\end{equation}
and the corresponding CR bound is saturable (as $\imag(\oG_1\oG_2)=0$). By a direct calculation one can see that the cost of joint estimation $\jntcr$ is minimized for the state $\ket{\psi}=\frac{1}{2}(\ket{1}+\ket{2}+\ket{3}+\ket{4})$ and that for this sate:
\begin{equation}
\label{examjnt}
\jntcr\limeq \frac{1}{k}\tr(F^{-1})=\frac{1}{k}\left(\frac{2}{(\alpha-\beta)^2}+\frac{2}{(\alpha+\beta)^2}\right).
\end{equation}
In order to calculate $\seppcr$, instead of performing a direct optimization given in \eqref{crsep}, we just use the fact that $\seppcr\geq\jntcr$ and show a particular transformation $A$ for which this bound is saturated. Let us choose as new parameters $\var_1'=\alpha\var_1+\beta\var_2$, $\var_2'=\beta\var_1+\alpha\var_2$, so \begin{equation}
A^{-1}=\begin{bmatrix}
\alpha & \beta\\
\beta & \alpha\\
\end{bmatrix}\Rightarrow
A=\frac{1}{\alpha^2-\beta^2}\begin{bmatrix}
\alpha & -\beta\\
-\beta & \alpha\\
\end{bmatrix}.
\end{equation}
After such a transformation the differences between extreme eigenvalues of new generators are both equal $\forall_i\lambda[A^T\bG_i]=1$, and the new parameters may be effectively measured with states $\frac{1}{\sqrt{2}}(\ket{1}+\ket{2})$, $\frac{1}{\sqrt{2}}(\ket{3}+\ket{4})$, which leads to:
\begin{equation}
\label{nonorto}
\seppcr\limeq \frac{1}{k}\left(\sum_{i=1}^p\sqrt{[A^2]_{ii}}\right)^2=\frac{1}{k}\left(\frac{2}{(\alpha-\beta)^2}+\frac{2}{(\alpha+\beta)^2}\right),
\end{equation}
which is indeed equal to \eqref{examjnt}. Finally, let us show that this results is not achievable, if one restricted just orthogonal transformations $O$. Let
 \begin{equation}
O^{-1}=
\begin{bmatrix}
\boldsymbol{o}_1, &\boldsymbol{o}_2
\end{bmatrix}^T
=
\begin{bmatrix}
\cos(\varphi) & -\sin(\varphi)\\
\sin(\varphi) & \cos(\varphi)\\
\end{bmatrix}.
\end{equation}
Then $[O^TO]_{ii}=1$ and $[O^{-1}F_i^{-1}O^{-1T}]_{ii}=\boldsymbol{o}_i^TF_i^{-1}\boldsymbol{o}_i$ so:
\begin{equation}
\label{orto}
{\seppcr}_{\t{,ortho}}\limeq \frac{1}{k}\min_{\varphi}\left(\sum_{k=1}^2\sqrt{\min_{\ket{\psi_k}}\boldsymbol{o}_i^T[F_i]^{-1}\boldsymbol{o}_i}\right)^2.
\end{equation}
Both \eqref{nonorto} and \eqref{orto} are compared in \figref{fig:A} for different ratios between $\alpha$ and $\beta$. One may see that for the ration around $1/2$ a significant advantage due to application of the non-orthogonal transformation may be observed.

\section{Relation between the optimal global group invariant cost and the local quadratic minimax cost in the limit of large $N$}
\label{App:cost}
In this appendix, we formalize the reasoning from section \ref{sec:locglob}.
\bigskip

\textbf{Theorem.} Consider a quantum channel $U_g$, $g \in G$, which corresponds to a unitary representation of a compact group $G$ in some Hilbert space $U_{g_1}U_{g_2}=U_{g_1g_2}$, and the cost function is invariant with respect to the action of the group $\forall_{g,\tilde g, h\in G} \cf(hg,h\tilde g)=\cf(g,\tilde g)$. Let $\bvar=[\var_1,...,\var_p]$ be a local parametrization around some $g_0\in G$. Let $G_\delta\subset G$ be the subset of $G$ containing all $g_\bvar$ such that $\forall_i \var_i\in[-\delta/2,+\delta/2]$. Then, for the  most general adaptive scheme, the local asymptotic minimax cost is the same as the global asymptotic minimax cost:
\begin{equation}
\label{theorem}
\begin{split}
\inf_{\{M_{N,\tilde g},\rho_{N, g}\}}\lim_{\delta\to 0}\lim_{N\to\infty}N^2 \sup_{g\in G_{\delta}}\int d\tilde g \tr(M_{N,\tilde g}\rho_{N,g}) \cf(g,\tilde g)
&=\inf_{\{M_{N,\tilde g},\rho_{N, g}\}}\lim_{N\to\infty}N^2 \sup_{g\in G}\int d\tilde g \tr(M_{N,\tilde g}\rho_{N,g})\mathcal \cf(g,\tilde g)
\end{split}
\end{equation}
Moreover, as obtaining optimal global asymptotic minimax cost was proven not to require adaptiveness~\cite{chiribella2008memory}, the above equation remains valid when restricting the RHS to parallel strategies. It also implies, that there is no asymptotic advantage in applying adaptive strategy also in the local case.

\textit{Proof.}
Let us introduce the notation for the minimax cost with a finite $\delta, N$:
\begin{equation}
\t{minimax}(G_\delta,N)=\inf_{M_{N,\tilde g},\rho_{N, g}}\sup_{g\in G_\delta}\int d\tilde g \tr(M_{\tilde g}\rho_{N,g})\cf(g,\tilde g).
\end{equation}
We would like to prove that in the limit of large $N$, the value of $\delta$ has no impact on the final cost. It is clear that:
\begin{equation}
\label{trivial}
\forall_{\delta}\t{minimax}(G_\delta,N)\leq \t{minimax}(G,N).
\end{equation}
To bound the cost for finite $\delta$ from below, we use the following construction~\cite{hayashi2011,gorecki2021multiple}.
Having at our disposal $N$ gates in total we  may at first perform $\sqrt{N}$ independent measurements, to find an approximated value $g_{\t{est}}$. Due to the central limit theorem, the probability, that $g\notin g_{\t{est}}G_{\delta}$ ($g_{\t{est}}G_{\delta}$ is the set $G_{\delta}$ shifted by the action of $g_{\t{est}}$) decreases exponentially $p_{\t{err}}(\sqrt{N})\propto e^{-\sqrt{N}}$. Next, we spend the remaining $N-\sqrt{N}$ gates to perform estimation around the point $g_{\t{est}}$. Since from the point of view of the initial problem
of estimating an unknown $g$ with $N$ gates, such a procedure might be suboptimal, we have:
\begin{equation}
\label{prsub}
\t{minimax}(G,N)\leq p_{\t{err}}(\sqrt{N})c_{\max}+(1-p_{\t{err}}(\sqrt{N}))\t{minimax}(g_{\t{est}}G_{\delta},N-\sqrt{N}),
\end{equation}
where $c_{\max}=\max_{g,\tilde g}\cf(g,\tilde g)$. Moreover, due to the symmetry of the whole problem, the RHS does not depend on
$g_{\t{est}}$. After application of $\lim_{N\to\infty}N^2\cdot$ to the both sites, and the use of $\lim_{N\to\infty}\frac{(N-\sqrt{N})^2}{N^2}=1$, we obtain:
\begin{equation}
\forall_{\delta}\lim_{N\to\infty}N^2\t{minimax}(G,N)\leq\lim_{N\to\infty}N^2(G_\delta,N-\sqrt{N})=\lim_{N\to\infty}N^2(G_\delta,N),
\end{equation}
which, together with \eqref{trivial} gives:
\begin{equation}
\forall_{\delta}\lim_{N\to\infty}N^2\t{minimax}(G_\delta,N)= \lim_{N\to\infty}N^2\t{minimax}(G,N).
\end{equation}
Finally, since in general for any functional family $\lim_{x\to x_0}\inf_yF_x(y)\leq \inf_y\lim_{x\to x_0}F_x(y)$, the LHS of \eqref{theorem} may be bounded from below by $\lim_{\delta\to 0}\lim_{N\to\infty}N^2\t{minimax}(G_\delta,N)=\lim_{N\to\infty}N^2\t{minimax}(G,N)$, while the RHS is exactly equal to $\lim_{N\to\infty}N^2\t{minimax}(G,N)$. This ends the proof. $\square$

\bigskip
It is worth to note, that if all the elements of the group representation commute $[U_{g_1},U_{g_2}]=0$, then in order to implement the strategy \eqref{prsub} one does not need adaptivness and the procedure may be performed within the parallel scheme. Indeed, in such a case there is a single state optimal for local measurements around an arbitrary point $g_{\t{est}}$, as rotating the state is equivalent to rotating the measurement in the opposite direction $\tr(M U_g U_{g_{\t{est}}^{-1}}\rho_0 U_{g_{\t{est}}^{-1}}^\dagger U_g^\dagger)=\tr(U_{g_{\t{est}}^{-1}}^\dagger MU_{g_{\t{est}}^{-1}} U_g \rho_0  U_g^\dagger)$---so the knowledge of the value of $g_{\t{est}}$ is not needed at the level of state preparation. This is, however, no longer true if $[U_{g_1},U_{g_2}]\neq 0$. Still, regardless of this sub-optimal strategy used in the proof, a fully optimal strategy does not require adaptiveness and may be performed within parallel scheme~\cite{chiribella2008memory}, so theorem remains valid even when restricted to parallel strategies.

\subsection*{Quadratic cost approximation}


As shown in~\cite[Section 9]{yang2019attaining} \cite[Section II-D]{hayashi2008asymptotic}, in local estimation, analyzed within many repetition scenario, even if one consider general cost function, in the limit $k\to\infty$ it may be well approximated by the quadratic term $\tr(W\cov)=(\tilde\bvar-\bvar)^TW(\tilde\bvar-\bvar)$ (where $W$ is the Hessian of the cost function). However, in principle, it is not so clear in the single repetition approach, as while in the many repetition scenario, due to the central limit theorem, all probabilities converge to the gaussian ones (with exponentially decreasing tails), in the single-shot case the strategy minimizing the cost may lead to much slower decreasing tails~\cite{Imai2009}.

More formally, let $W$ be the Hessian of the cost function around $g_0$, i.e. $W_{ij}=\partial_{\var_i}\partial_{\var_j}\cf(g_0,g_\bvar)$. Then for the particular strategy $\{M_{N,\tilde \bvar},\rho_{N,\bvar}\}$ which is known to lead to the cost $\propto\frac{1}{N^2}$ it is not clear if in the point $\bvar=[0,...,0]$:
\begin{equation}
\lim_{N\to\infty}N^2\int d\tilde \bvar \tr(M_{N,\tilde \bvar}\rho_{N,0})\cf(g_0,g_{\tilde\bvar})\overset{?}{=}\lim_{N\to\infty}N^2\int d \tilde \bvar \tr(M_{N,\tilde \bvar}\rho_{N,\bvar})\sum_{ij}W_{ij}\tilde\var_i\tilde\var_j
\end{equation}
as in principle it may happen that $\tr(M_{N,\tilde \bvar}\rho_{N,0})$ decrease like $\propto \frac{1}{N^2|\bvar|^2}$ for large $\bvar$ (making also higher derivatives of $\cf(g_0,g_{\tilde\bvar})$ not negligible).  Below we prove, that for the strategy minimizing LHS of above, such a tail always decreases fast enough (because if it were otherwise, one could use some small part of gates $\sim\sqrt{N}$ to cut this tail).

\bigskip

\textbf{Theorem.} For the optimal (both adaptive or parallel) global strategy $\{M^{\t{opt}}_{N,\tilde g},\rho^{\t{opt}}_{N, g}\}$,
invariant under the group action, for the cost $\cf(g_{0},g_{\tilde\bvar})=\sum_{ij}W_{ij}\tilde\var_i\tilde\var_j+o\left(\tilde\var_i\tilde\var_j\right)$ (where $W>0$):
\begin{equation}
\lim_{N\to\infty}N^2\int d\tilde g \tr(M^{\t{opt}}_{N,\tilde g}\rho^{\t{opt}}_{N,g_0})\cf(g_0,\tilde g)=\lim_{N\to\infty}N^2\int d \tilde \bvar \tr(M^{\t{opt}}_{N,g_{\tilde \bvar}}\rho^{\t{opt}}_{N,g_0})\sum_{ij}W_{ij}\tilde\var_i\tilde\var_j
\end{equation}

\textit{Proof.} For any $g$ we split the mean cost integral  into two parts:
\begin{equation}
\forall_{g\in G}\int_G d \tilde g \tr(M^{\t{opt}}_{N,\tilde g}\rho^{\t{opt}}_{N,g})\cf(g,\tilde g)=
\underbrace{\int_{G_{\delta}} d \tilde g \tr(M^{\t{opt}}_{N,\tilde g}\rho^{\t{opt}}_{N,g})\cf(g,\tilde g)}_{C_1(N,\delta)}
+
\underbrace{\int_{G\setminus G_\delta} d \tilde g \tr(M^{\t{opt}}_{N,\tilde g}\rho^{\t{opt}}_{N,g})\cf(g,\tilde g)}_{C_2(N,\delta)}.
\end{equation}
Next, we prove by contradiction that $\forall_{\delta>0}\lim_{N\to\infty}N^2C_2(\delta,N)=0$.

\begin{figure}[t]
\includegraphics[width=0.70 \textwidth]{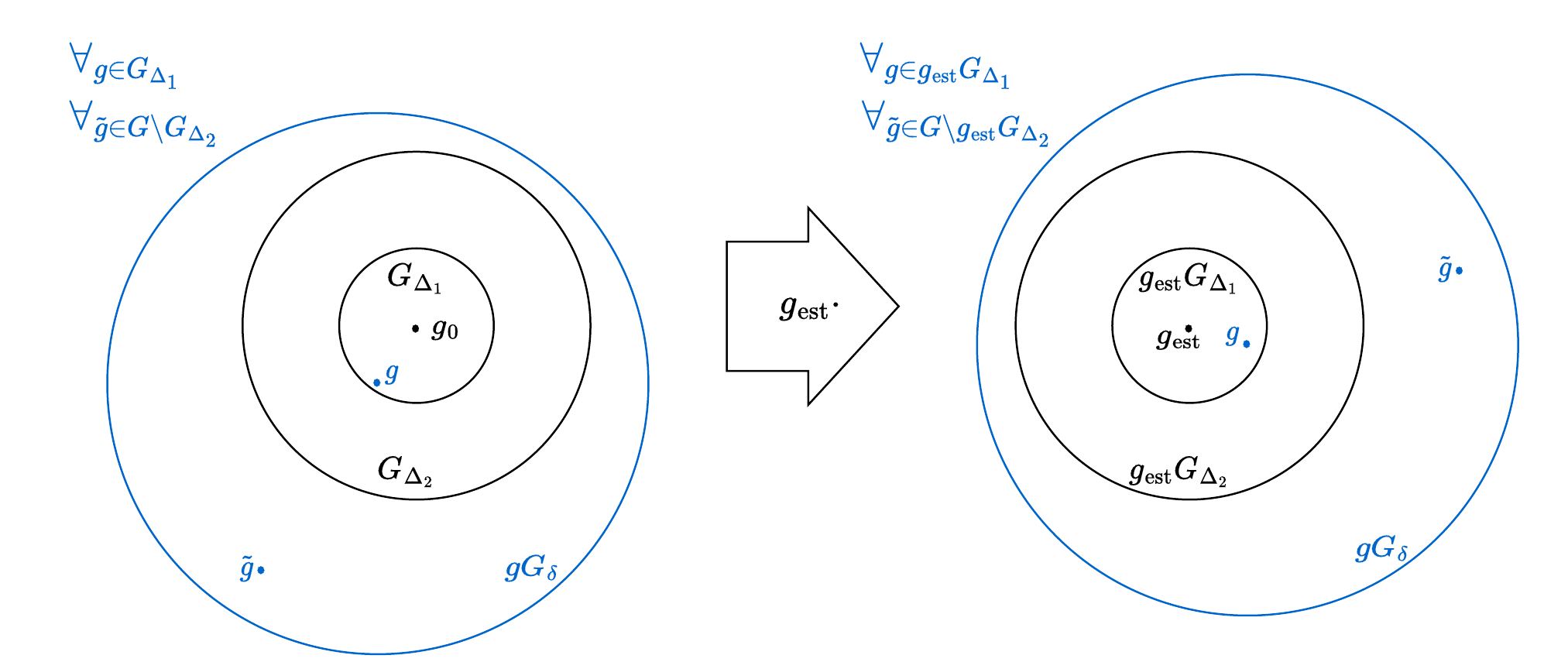}
\caption{(Color online) Graphical illustration of the proof. One spend at first $\sqrt{N}$ gates to find $g_{\t{est}}$ such that probability of finding true value of $g$ outside of $g_{\t{est}}G_{\Delta_1}$ is negligible. Then each time where the second part of the measurement point $\tilde g\notin g_{\t{est}}G_{\Delta_2}$, it is more efficient to set $\tilde g=g_{\t{est}}$.
}
\label{fig:G}
\end{figure}

Assume, that $\lim_{N\to\infty}N^2C_2(\delta,N)>0$. For a given $\delta$ let us choose finite neighborhoods of $g_0$, namely $G_{\Delta_1}$, $G_{\Delta_2}$, such that $G_{\Delta_1}\subset G_{\Delta_2}$ and $\forall_{g\in G_{\Delta_1}}G_{\Delta_2}\subset gG_{\delta}$, satisfying $\max_{g\in G_{\Delta_1}}\cf(g_0,g)< \min_{g\in G_{\Delta_1},\tilde g\in G\setminus G_{\Delta_2}}\cf (g,\tilde g)$ (see left part of \figref{fig:G}).

Next, similarly as in the previous proof, one may at first spend $\sqrt{N}$ gates to find $g_{\t{est}}$, such that the probability that the true value of $g$ lays outside of $g_{\t{est}}G_{\Delta_1}$ decreases exponentially $p_{\t{err}}(\sqrt{N})\propto e^{-\sqrt{N}}$. Then one uses the remaining $N-\sqrt{N}$ gates to perform the mentioned $M^{\t{opt}}_{N-\sqrt{N},\tilde g}\rho^{\t{opt}}_{N-\sqrt{N},g}$ with the following correction: each time when $\tilde g$ points outside of $g_{\t{est}}G_{\Delta_2}$, one is forced to estimate $\tilde g=g_{\t{est}}$. The total cost for such a constructed strategy may be bounded from above:
\begin{equation}
\leq p_{\t{err}}(\sqrt{N})c_{\max}+(1-p_{\t{err}}(\sqrt{N}))(C_1(N-\sqrt{N},\delta)+C_\Delta (N-\sqrt{N},\delta))
\end{equation}
where $c_{\max}=\max_{g,\tilde g\in G}\cf(g,\tilde g)$ and
\begin{equation}
C_\Delta(N-\sqrt{N},\delta)=\left(\int_{G\setminus G_\delta} d \tilde g \tr(M^{\t{opt}}_{N-\sqrt{N},\tilde g}\rho^{\t{opt}}_{N-\sqrt{N},g})\right)\cdot\max_{g\in G_{\Delta_1}}\cf(g_0,g).
\end{equation}
Since
\begin{equation}
C_2(N-\sqrt{N},\delta)\geq\left(\int_{G\setminus G_\delta} d \tilde g \tr(M^{\t{opt}}_{N-\sqrt{N},\tilde g}\rho^{\t{opt}}_{N-\sqrt{N},g})\right)\cdot\min_{g\in G_{\Delta_1},\tilde g\in G\setminus G_{\Delta_2}}\cf (g,\tilde g)
\end{equation}
we have $\lim_{N\to\infty}N^2\left(C_2(N-\sqrt{N},\delta)-C_\Delta(N-\sqrt{N},\delta)\right)>0$. It means that cutting the tail indeed decreases the cost, which leads to a contradiction with the assumption about the optimality of $\{M^{\t{opt}}_{N,\tilde g},\rho^{\t{opt}}_{N, g}\}$. Now, as from $\mathcal C(g_{0},g_{\tilde\bvar})=W_{ij}\tilde\var_i\tilde\var_j+o(\tilde\var_i\tilde\var_j)$ we have
$\forall_{\epsilon}\exists_\delta\forall_{-\delta/2\leq \tilde\var_i\leq +\delta/2} |\mathcal C(g_0,g_{\tilde\bvar})-W_{ij}\var_i\var_j|\leq \epsilon$, the statement is proven.  $\square$

\bigskip

From the above, if $W_{ij}=\delta_{ij}$, then for the optimal strategies the cost is equivalent to $\Delta^2\tilde\bvar$ discussed in the main paper. Below we show, that this is indeed the case for examples discussed in section \ref{sec:commuting}.

In~\cite{chiribella2004} the authors considered the problem of transmitting a reference frame by sending $N$ spin-$1/2$ atoms, which is equivalent to the estimation of a completely unknown element of $\t{SU}(2)$. As the figure of merit they chose the error function given as:
\begin{equation}
e(g,\tilde g)=6-2\tr(U^{(1)}_g U^{(1)\dagger}_{\tilde g})
\end{equation}
where $U^{(1)}_g$ is the rotation matrix of the spin-1 particle. By a direct calculation
\begin{equation}
\tr(U^{(1)}_{g_{0}}U^{(1)\dagger}_{g_{\tilde\bvar}})=1+2\cos(|\bvar|),
\end{equation}
so
\begin{equation}
e(g_0,g_{\tilde\bvar})=4(1-\cos(|\tilde\bvar|))=8\sin^2(|\tilde\bvar|/2).
\end{equation}
Therefore indeed we have:
\begin{equation}
\cf(g_0,g_{\tilde\bvar})=\frac{1}{2}e(0,g_{\tilde\bvar})=4\sin^2(|\tilde\bvar|/2)=\tilde\bvar^2+o(\tilde\var_i\tilde\var_j).
\end{equation}

In~\cite{bagan2000,bagan2001} the authors considered the problem of transmitting a unit vector $\vec{n}$ by sending $N$ spin-$1/2$ atoms. After introducing parametrization $\vec{n}=[\cos(\vartheta)\cos(\varphi),\cos(\vartheta)\sin(\varphi),\sin(\vartheta)]$ with $\varphi=\sqrt{\var_1^2+\var_2^2}$ and $\vartheta=\arctan(\var_1/\sqrt{\var_1^2+\var_2^2})$, the problem is equivalent to the estimation of the channel $U_{g_{\bvar}}=e^{i(\var_1\sigma_x/2+\var_2\sigma_y/2)}$. As the figure of merit they chose fidelity:
\begin{equation}
F(\vec{n},\vec{\tilde n})=(1\pm \vec{n}\cdot{\vec{\tilde n}})/2,
\end{equation}
for which we have:
\begin{equation}
F(\vec{n}_{0},\vec{n}_{\tilde\bvar})=(1+\cos(|\tilde\bvar|))/2=1-\sin^2(|\tilde\bvar|/2),
\end{equation}
so indeed:
\begin{equation}
\cf(\vec{n}_{0},\vec{n}_{\tilde\bvar})=4(1-F(\vec{n}_{0},\vec{n}_{\tilde\bvar}))=4\sin^2(|\tilde\bvar|/2)=\tilde\bvar^2+o(\tilde\var_i\tilde\var_j).
\end{equation}

\section{Derivation of the bound and an exemplary state for estimation of magnetic field in $p$ points of space}
\label{App:airy}

\subsection{Analitical solutions for $p=1,2$}
For $p=1$ the simplex $\sum_{i=1}^p|\mu_i|\leq 1/2$ is simply the line $\mu_1\in[-1/2,+1/2]$, so the optimal solution is $f(\mu_1)=\sqrt{2}\cos(\pi\mu_1)$ with the cost $\pi^2/N^2$. For $p=2$ the simplex $\sum_{i=1}^p|\mu_i|\leq 1/2$ takes the form of a square with side $\sqrt{2}/2$ rotated by the angle $45^\circ$ relative to the coordinate axes, which allows for an effective coordinates separation; therefore the solution is $f(\mu_1,\mu_2)=2\cos(\sqrt{2}\pi(\mu_1+\mu_2)/\sqrt{2})\cos(\sqrt{2}\pi(\mu_1-\mu_2)/\sqrt{2})$ with the corresponding cost $\jntmm=2\times (\sqrt 2)^2\pi^2/N^2=4\pi^2/N^2$.

\subsection{Derivation of the bound}
For the problem \eqref{singmin}:
\begin{align}
\jntmm\geq \frac{p}{N^2}\min_g\int_{0}^{+\infty}d\mu g^*(\mu)\left(-\frac{\partial^2}{\partial^2\mu}\right)g(\mu),\quad \t{with}\quad &\frac{\partial g}{\partial\mu}\Big|_{\mu=0}=0,\label{main}\\
&\int_{0}^{+\infty}d\mu |g(\mu)|^2=1,\label{conone}\\
&\int_{0}^{+\infty}d\mu |g(\mu)|^2|\mu|=\frac{1}{2p},\label{contwo}
\end{align}
the solution may be found using the standard Lagrange multiplier method,
\begin{equation}
-\frac{\partial^2}{\partial \mu^2}g(\mu)+g(\mu)(\lambda_1+\mu\lambda_2)=0 \Rightarrow g(\mu)\propto \t{Ai}\left(
\lambda_2^{1/3}(\lambda_1\lambda_2^{-1/3}+\mu)\right),
\end{equation}
where $\t{Ai}(\cdot)$ is the Airy function of the first kind. The condition $\frac{\partial g}{\partial\mu}\Big|_{\mu=0}=0$ implies
$\lambda_1\lambda_2^{-1/3}=A'_0\approx -1.019$, where $A'_0$ is the first zero of derivative of $\t{Ai}(\cdot)$.
From \eqref{main} and \eqref{conone} we have:
\begin{equation}
\jntmm\geq \frac{p}{N^2}\frac{\int_{0}^{+\infty}d\mu |\partial_{\mu}\t{Ai}(\lambda_2^{1/3}(A_0'+\mu))|^2}{\int_{0}^{+\infty}d\mu |\t{Ai}(\lambda_2^{1/3}(A_0'+\mu))|^2}=\frac{\int_{0}^{+\infty}d\mu |\partial_{\mu}\t{Ai}((A_0'+\mu))|^2}{\int_{0}^{+\infty}d\mu |\t{Ai}((A_0'+\mu))|^2}\cdot\lambda_2^{2/3}.
\end{equation}
To get the value of $\lambda_2$ we use \eqref{contwo} and \eqref{conone}:
\begin{equation}
\frac{1}{2p}=
\frac{\int_{0}^{+\infty}d\mu |\t{Ai}(\lambda_2^{1/3}(A_0'+\mu))|^2\mu}{\int_{0}^{+\infty}d\mu |\t{Ai}(\lambda_2^{1/3}(A_0'+\mu))|^2}=\frac{\int_{0}^{+\infty}d\mu |\t{Ai}(A_0'+\mu)|^2\mu}{\int_{0}^{+\infty}d\mu |\t{Ai}(A_0'+\mu)|^2}\cdot \lambda_2^{-1/3},
\end{equation}
so finally:
\begin{equation}
\jntmm\geq \frac{p^3}{N^2}\frac{4(\int_{0}^{+\infty}d\mu |\partial_{\mu}\t{Ai}((A_0'+\mu))|^2)(\int_{0}^{+\infty}d\mu |\t{Ai}((A_0'+\mu))|^2\mu)^2}{(\int_{0}^{+\infty}d\mu |\t{Ai}((A_0'+\mu))|^2)^3}\approx \frac{0.63p^3}{N^2}.
\end{equation}

\subsection{Exemplary state}

Finally, let us present a suboptimal, but an explicit analytical solution of the initial problem \eqref{tominimize}, which shows a significant advantage compared with the optimal SEP protocol. Namely, we choose the largest possible $p$-dimensional ball inside the simplex $\sum_i|\mu_i|\leq 1/2$ and then take as the $f(\boldsymbol \mu)$ the function which minimizes the kinetic energy inside this ball with a boundary condition $f(\boldsymbol \mu)=0$ on the border and outside of the ball. The Laplacian for spherical coordinated is given as:
\begin{equation}
\Delta f=\frac{\partial^2 f}{\partial r^2}+\frac{p-1}{r}\frac{\partial f}{\partial r}+\t{angular part},
\end{equation}
where the exact form of the angular part is irrelevant for the discussion. The corresponding eigenstates are the ones of the form:
\begin{equation}
f(r)\propto r^{(2-p)/2} J_{p/2-1}(\sqrt{E}r)\Rightarrow -\Delta f(r)=E f(r),
\end{equation}
where $J_{\alpha}(\cdot)$ is the Bessel function of the first kind. As the radius of the biggest ball inside the simplex $R$ satisfies:
\begin{equation}
R^2=\sum_{i=1}^p (1/2p)^2=\frac{1}{4p},
\end{equation}
and taking into account the boundary condition $f(R)=0$ we have:
\begin{equation}
\sqrt{E}\frac{1}{2\sqrt{p}}=j_{p/2-1,1}\Rightarrow E=p (2j_{p/2-1,1})^2,
\end{equation}
where $j_{p/2-1,1}$ is the zero of the Bessel function $J_{p/2-1}(x)$. Since for large $p$ we have $j_{p/2-1,1}\approx p/2$, we get:
\begin{equation}
E\approx p^3,
\end{equation}
which yields the joint cost $\frac{p^3}{N^2}$.

\end{document}